\documentclass[12pt]{article}
\usepackage{amsmath,amssymb,bm,epsfig}
\renewcommand{\thefootnote}{\fnsymbol{footnote}}

\newcommand{\1}{\mbox{1}\hspace{-0.25em}\mbox{l}}

\begin{document}

\title{
\begin{flushright}
\begin{minipage}{0.2\linewidth}
\normalsize
%arXiv:YYMM.NNNN \\
WU-HEP-15-12  \\*[50pt]
\end{minipage}
\end{flushright}
{\Large \bf 
Wavefunctions on magnetized branes in the conifold
\\*[20pt]}}

\author{Hiroyuki~Abe$^{1,}$\footnote{
E-mail address: abe@waseda.jp}, \ 
Akane~Oikawa$^{1,}$\footnote{
E-mail address: a.oikawa@aoni.waseda.jp}
\ and \ 
Hajime~Otsuka$^{1,}$\footnote{
E-mail address: hajime.13.gologo@akane.waseda.jp
}\\*[20pt]
$^1${\it \normalsize 
Department of Physics, Waseda University, 
Tokyo 169-8555, Japan} \\*[50pt]}

\date{
\centerline{\small \bf Abstract}
\begin{minipage}{0.9\linewidth}
\medskip 
\medskip 
\small
We study wavefunctions on D$7$-branes 
with magnetic fluxes in the conifold. 
Since some supersymmetric embeddings of D-branes on the $AdS_5\times T^{1,1}$ geometry are known,
we consider one of the embeddings, especially the 
spacetime filling D$7$-branes in which (a part of) the standard 
model is expected to be realized. 
The explicit form of induced metric on the D$7$-branes 
allows us to solve the Laplace and Dirac equations to evaluate matter wavefunctions in extra dimensions analytically.
We find that the zero-mode wavefunctions can be 
localized depending on the configuration of magnetic fluxes on D$7$-branes, 
and show some phenomenological aspects. 
\end{minipage}
}

\begin{titlepage}
\maketitle
\thispagestyle{empty}
\clearpage
\tableofcontents
\thispagestyle{empty}
\end{titlepage}

\renewcommand{\thefootnote}{\arabic{footnote}}
\setcounter{footnote}{0}

\section{Introduction}
One of the aim in the subject of string phenomenology 
is to find out the true vacuum of string theory which contains the standard 
model (SM) as the effective theory. 
Moreover, in the framework of string models, the masses and 
mixing angles of the four-dimensional ($4$D) chiral matter fields would be 
dynamically generated, although they are fundamental parameters 
in the SM. 
It is then expected that the masses of quarks and leptons 
in the SM are clues to find out the realistic string models 
by comparing the theoretical values with the observed ones. 

Since the coupling constants in the $4$D effective theory, such as Yukawa couplings, are determined by the overlap integrals of 
wavefunctions in the internal manifold such as 
the Calabi-Yau (CY) manifold, 
they are affected by the geometrical 
structures behind the $4$D spacetime. 
Therefore, we concentrate on the behaviors of 
matter wavefunctions on some curved backgrounds in this paper. 
Such an approach to construct string models on 
various geometrical backgrounds would give us a 
guiding principle to determine the internal geometry, 
especially the CY manifold. 

In the type IIB string theory, the non-Abelian gauge groups appear from D-branes whose low-energy effective action is described as the supersymmetric Yang-Mills (SYM) theory. 
As one of the simplest setup, one usually considers the toroidal background.
For example, the authors of Ref.~\cite{Cremades:2004wa} derived the wavefunctions for chiral matter zero-modes and identified their degeneracy by employing the 
Yang-Mills fluxes on D-branes, that is, magnetized D-branes. 

It is then argued that the number of zero-modes 
are interpreted as that of generations for the chiral matter fields, 
which is determined by the magnetic flux configurations. 
Because of the quasi-localized profiles of wavefunctions, one can obtain, e.g., 
a hierarchical structure of the Yukawa couplings. 
Recently, it has been shown that the obtained Yukawa couplings possess some discrete flavor symmetries~\cite{Abe:2009vi}, 
especially in the ten-dimensional ($10$D) SYM theory compactified on three factorized tori, 
$T^2 \times T^2 \times T^2$, with magnetic fluxes~\cite{Abe:2009vi,Abe:2012fj} and 
it yields some semi-realistic patterns of quark and lepton masses and mixing angles with 
supersymmetric~\cite{Abe:2009vi,Abe:2012fj} and non-supersymmetric~\cite{Abe:2014vza} flux configurations. 

On the other hand, there are some studies for the 
local magnetized D$7$-brane models on the curved 
backgrounds such as 
$\mathbb{P}^1$, $\mathbb{P}^1 \times \mathbb{P}^1$ and 
$\mathbb{P}^2$~\cite{Conlon:2008qi} and the authors 
of Ref.~\cite{Conlon:2008qi}  
show the explicit matter wavefunctions. 
So far, the local models are considered to be embedded in 
some global CY threefold. 
Since the analytical metric of any explicit CY 
threefold is not known at the moment, it is a challenging issue to compute matter wavefunctions analytically 
in a global model. 
In this paper, we try to evaluate them in a certain 
local model on the conifold which may be embedded into some class of CY manifold.

In order to obtain the local description of chiral matters in an explicit 
CY manifold, we adopt the Klebanov-Witten model in 
which the geometry is described by the $AdS_5 \times T^{1,1}$ 
due to the stack of a large number of D$3$-branes 
placed at the tip of conifold~\cite{Klebanov:1998hh}. 
These conical singularities are ubiquitous in string theory. 
For example, when a large number of D$3$-branes are placed at the same point 
in the internal space, such conical singularities appear due to the 
backreaction~\cite{Verlinde:1999fy}. Also, a statistical analysis 
has shown such conical singularities are common in 
the landscape of string theory~\cite{Hebecker:2006bn}. 
As a local model included in the Klebanov-Witten background, 
we consider the probe D$7$-branes which 
wrap the certain cycles in the conifold, and 
study the matter wavefunctions living on them. 
When the D-branes wrap some cycles, 
the stability of them can be verified by the existence of the kappa-symmetry~\cite{Aganagic:1996pe}, 
which is a useful probe to search for the local calibrated cycles. 
Since the kappa-symmetry is accompanied by the local supersymmetry 
on the geometrical background, the wrapped D-branes 
preserve at least ${\cal N}=1$ supersymmetry. 
These techniques are developed especially on the 
$AdS_5 \times T^{1,1}$ background in Ref.~\cite{Arean:2004mm}. 

By employing them, we will find the phenomenologically 
attractive D-brane models, that is, spacetime filling 
ones, although 
the different D-brane configurations from those we adopt have attracted lots of attention so far 
in the light of AdS/CFT correspondence. (See for more details, e.g., 
Refs.~\cite{Arean:2004mm,Canoura:2005uz}.) 
Therefore, in this paper, we consider spacetime filling D$7$-branes 
which wrap the supersymmetric four-cycles on 
$AdS_5 \times T^{1,1}$ and then the magnetic fluxes 
are inserted from the phenomenological point of view. 
Then, the Dirac and Laplace equations in terms of the 
induced metric on D$7$-brane are analytically solved, which 
lead to localized chiral zero-modes with certain degeneracies, yielding some hierarchical structures in the low energy effective theory.
%As discussed in Ref.~\cite{Cascales:2003wn}, the semi-realistic spectrum of D$3$-branes 
%are realized at the orbifold singularity at the bottom of a 
%warped throat such as the conifold.
%These D-brane configurations are also interesting from the 
%cosmological point of view. On the warped these backgrounds, it have been 
%pointed out the possibility of brane inflation on the warped deformed conifold~\cite{Kachru:2003sx} 
%as well as the natural inflation on the warped resolved conifold~\cite{Kenton:2014gma}, 
%respectively. 

This paper is organized as follows. 
In Sec.~\ref{sec:2}, we review the Klebanov-Witten 
model and the kappa-symmetry. 
The decomposition of fields on D$7$-branes is also explained in this section for a latter convenience.
Then, the analytic solutions are derived in Sec.~\ref{sec:3} for 
the Laplace and Dirac equations on D$7$-branes 
with magnetic fluxes. 
In Sec.~\ref{sec:AdS5}, we show some phenomenological aspects of localized zero-modes in some effective descriptions with five and four spacetime dimensions. 
Finally we conclude in Sec.~\ref{sec:conclu}. 
The kappa-symmetry condition is summarized in Appendix~\ref{sec:A}, and the explicit forms of spin connections relevant to our analysis are shown in Appendix~\ref{sec:B}. 

\section{Supersymmetric brane probes on the conifold}
\label{sec:2}
In this section, we briefly review the supersymmetric 
embedding of D$7$-brane, which wrap a certain four-cycle in 
compact Calabi-Yau three-fold in order to study 
the matter field wavefunction living on the D$7$-brane. 
As an illustrative model, we focus on the Klebanov-Witten 
background caused by the stack of a large number of D$3$-branes, and introduce magnetized D$7$-branes there, assuming the backreaction to the background spacetime is negligible. 
Although the original Klebanov-Witten model is 
established on the non-compact Calabi-Yau manifold, i.e., 
the conifold, we assume that the conifold is locally 
described in a certain limit of some global Calabi-Yau 
manifold throughout this paper. 
The supersymmetric embedding of D$7$-brane is ensured 
from the analysis of a local fermionic symmetry called 
kappa-symmetry which implies that the D$7$-brane has 
at least ${\cal N}=1$ supersymmetry. 
Finally, we obtain the eight-dimensional (8D) SYM action as a low energy 
effective limit of the Dirac-Born-Infeld action for the D$7$-brane. 

\subsection{Klebanov-Witten model}
Before discussing the detail of the Klebanov-Witten model, 
we show the geometry of conifold which is defined as the complex 
three-dimensional hypersurface in $\mathbb{C}^4$,
\begin{eqnarray}
z_1z_2-z_3z_4 =0,
\label{eq:hol}
\end{eqnarray}
in terms of the four holomorphic coordinates $z_a$, $a=1,2,3,4$ 
on $\mathbb{C}^4$. 
From the above defining equation of the conifold, 
there is a conical singularity at the origin. 
It is well known that the metric of the conifold is described as a 
cone metric of $T^{1,1}$, which is the one of five-dimensional 
Sasaki-Einstein manifolds with cohomogenety one, and 
has homogeneous metric on $S^2 \times S^3$. 
The explicit form of a cone metric of $T^{1,1}$ 
is written as
\begin{align}
&ds^2=dr^2 + r^2 ds^2_{T^{1,1}},
\nonumber\\
&ds^2_{T^{1,1}}=\frac{1}{6} \sum_{i=1}^2 (d\theta_i^2 +\sin^2 \theta_i d\phi_i^2)
+\frac{1}{9} \left( d\psi +\sum_{i=1}^2 \cos\theta_i d\phi_i \right)^2,
\label{eq:T11}
\end{align}
where $r$ is the coordinate of $AdS_5$, while $\theta_i, \phi_i$ and $\psi$ 
are the coordinates of $T^{1,1}$ with $0\leq \theta_i<\pi$, $0\leq \phi_i <2\pi$ and $0\leq \psi <4\pi$. 
Since the cone metric is Ricci-flat and K\"{a}hler as can be 
seen in the definition of the Sasaki-Einstein manifolds, 
the conifold is a non-compact Calabi-Yau manifold. 
As mentioned before, we assume the conifold is a local 
description of certain Calabi-Yau threefold.

We proceed to review the type IIB supergravity solution, Klebanov-Witten model~\cite{Klebanov:1998hh}.
In this model, one starts with a stack of $N_c$ D$3$-branes at the tip of conifold with $g_sN_c\gg$1, where $g_s$ is the string coupling and 
$N_c$ is the number of D$3$-branes.
It causes the warped geometry near the conical singularity, and 
then the ten-dimensional spacetime becomes $AdS_5 \times T^{1,1}$ around the conical singularity supported 
by the Ramond-Ramond self-dual five-form flux $F^{(5)}$. 
In the near-horizon limit, such supergravity solution is written as, 
\begin{align}
ds_{10}^2 &=h(r)^{-1/2} d^2x_{1,3} + h(r)^{1/2} (dr^2 +r^2 ds_{T^{1,1}}^2), \nonumber\\
h(r)&=\frac{L^4}{r^4}, \nonumber\\
g_s F^{(5)} &=d^4 x \wedge dh^{-1} + \rm{Hodge\ dual}, \nonumber\\
L^4 &=\frac{27}{4} \pi g_s N_c \alpha'^2,
\label{eq:horizon}
\end{align}
where $\alpha^{\prime}$ is the regge slope, $dx^2_{1,3}$ is the line-element in the four-dimensional spacetime, $h(r)$ 
is the warp factor given by the backreactions of D$3$-branes. 
As discussed in Sec.~\ref{subsec:global}, a construction method for global Calabi-Yau is known and one includes the situation that the local conifold can be glued to the global compact Calabi-Yau 
manifold in the large radius limit of $r$.
 
\subsection{Kappa-symmetry for $AdS_5 \times T^{1,1}$}
In this section, we focus on the spacetime filling D$7$-brane wrapping 
the non-trivial four-cycle in the conifold\footnote{We can also consider the spacetime filling 
D$5$-brane embedded in the conifold, where the induced metric on the D$5$-brane is described by 
the projective space. In this case,  wavefunction as well as the Yukawa couplings are obtained in 
Ref.~\cite{Conlon:2008qi}.} and their worldvolume coordinates 
are denoted by
\begin{eqnarray}
\xi^\mu =(x^0, x^1, x^2, x^3 ,\theta_1, \phi_1,\theta_2,\phi_2),
\end{eqnarray}
where $\theta_i$, $\phi_i$ with $i=1,2$ are the coordinates of $T^{1,1}$ given in Eq.~(\ref{eq:T11}). 
In general, BPS configurations of the brane probe 
must satisfy the following condition for a worldvolume kappa-symmetry,
\begin{eqnarray}
\Gamma_\kappa \epsilon = \epsilon,
\label{eq:kappa}
\end{eqnarray}
where $\Gamma_\kappa$ is the kappa-symmetry matrix,
\begin{eqnarray}
\Gamma_\kappa =-\frac{i}{8!\sqrt{-g}} \epsilon^{\mu_1 \cdots \mu_8}\gamma_{\mu_1 \cdots \mu_8},
\end{eqnarray}
in the case of D$7$-brane, 
$\epsilon$ denotes Killing spinor for Klebanov-Witten background (2.4) and $\gamma_{\mu_1 \cdots \mu_8}$ is the antisymmetrized product of the 
gamma matrices pull-backed into the worldvolume of D$7$-brane. 
The explicit form of Killing spinors and kappa-symmetry matrix in arbitrary dimensions are summarized in Appendix~\ref{sec:A}. 
The kappa-symmetry condition ensures the invariance under the 
local supersymmetry transformations for dilatino and gravitino. 
As discussed in Ref.~\cite{Arean:2004mm}, the Eq.~(\ref{eq:kappa}) is solved in such a way that the 
transverse position of the D$7$-brane expressed by $r$ and $\psi$ 
changes depending on its worldvolume coordinates $\theta_i$ and $\phi_i$ 
with $i=1,2$ as
\begin{eqnarray}
r^3 &=&\frac{c^2}{\left( \sin \frac{\theta_1}{2} \right)^{n_1+1} \left( \cos \frac{\theta_1}{2} \right)^{1-n_1} \left( \sin \frac{\theta_2}{2} \right)^{n_2+1} \left( \cos \frac{\theta_2}{2} \right)^{1-n_2}},
\nonumber\\
\psi &=& n_1 \phi_1 + n_2 \phi_2 + \rm{const.},
\label{eq:rpsi}
\end{eqnarray}
where $n_1$ and $n_2$ are integers and $c$ is a constant, 
which determine the allowed region in the direction of $AdS_5$~\cite{Arean:2004mm}. 
In Eq.~(\ref{eq:rpsi}), $r$ is the specific function of the 
angles $\theta_i$ implying a non-zero minimal value of $r$ for $|n_i|\leq 1$ with $i=1,2$. 
On the other hand, when $\theta_i\rightarrow 0$ and/or $\pi$, the radial direction $r$ 
diverges and the D$7$-brane wrapping such four-cycle extends over 
the non-compact direction of the conifold. 

In general, the holomorphic coordinates $z_a$ for $a=1,2,3,4$ 
appeared in Eq.~(\ref{eq:hol}) can be written in terms of the worldvolume coordinates as
\begin{eqnarray}
z_1=r^{3/2} e^{\frac{i}{2} (\psi -\phi_1 -\phi_2)} \sin \frac{\theta_1}{2} \sin \frac{\theta_2}{2} ,\hspace{1cm} z_2=r^{3/2} e^{\frac{i}{2} (\psi +\phi_1 +\phi_2)} \cos \frac{\theta_1}{2} \cos \frac{\theta_2}{2}, \nonumber\\
z_3=r^{3/2} e^{\frac{i}{2} (\psi +\phi_1 -\phi_2)} \cos \frac{\theta_1}{2} \sin \frac{\theta_2}{2} ,\hspace{1cm} z_4=r^{3/2} e^{\frac{i}{2} (\psi -\phi_1 +\phi_2)} \sin \frac{\theta_1}{2} \cos \frac{\theta_2}{2}.
\label{eq:z}
\end{eqnarray}
Therefore the induced metric of the D$7$-brane,
\begin{align}
ds_{D7}^2 =\frac{1}{6}\sum_{i=1}^2 (d\theta_i^2 +\sin^2 \theta_i d\phi_i^2) &+
\frac{1}{9}(C_1(\theta_1)d\theta_1 +C_2(\theta_2)d\theta_2 )^2 \nonumber\\
&+\frac{1}{9}(C_1(\theta_1)\sin\theta_1 d\phi_1 +C_2(\theta_2)\sin\theta_2 d\phi_2 )^2,
\label{eq:D7metric}
\end{align}
where
\begin{eqnarray}
C_i(\theta_i)\equiv \frac{n_i +\cos\theta_i}{\sin\theta_i} \quad (i=1,2),
\end{eqnarray}
is rewritten in terms of the holomorphic coordinates $z_a$, that will be explicitly shown later in 
Eq.~(\ref{eq:D7metricz3z4}) for $n_1=n_2=1$.

\subsection{Mode expansions and dimensional reduction}
\label{subsec:red}
In this subsection, we describe mode expansions of scalars, spinors and vectors with respect to the 
internal coordinates following Ref.~\cite{Conlon:2008qi}. 
We first consider the $10$D $U(N)$ SYM as a 
low energy effective theory of D$9$-branes. 
The Lagrangian density is written as
\begin{eqnarray}
\mathcal{L}_{10D}=-\frac{1}{4g^2} {\rm Tr} \{ F^{MN}F_{MN}\}+\frac{i}{2g^2} {\rm Tr} \{\bar{\lambda}\Gamma^M D_M \lambda \},
\label{eq:10DSYM}
\end{eqnarray}
where $\lambda$ is the Majorana-Weyl spinor, $g$ is the gauge coupling, $M,N=0,\ldots ,9$ and the trace is denoted for the adjoint representation
of $U(N)$ gauge group. The field strength and covariant derivative 
are defined as
\begin{eqnarray}
F_{MN} &=&\partial_M A_N -\partial_N A_M -i [A_M,A_N],
\nonumber\\
D_M \lambda &=&\partial_M \lambda -i[A_M ,\lambda].
\label{eq:covariantD}
\end{eqnarray}
We can derive the $8$D SYM action for the spacetime filling D$7$-brane, reduced from the $10$D one. 
The dimensional reduction from $10$D to $8$D is achieved by integrating the functions depend on transverse coordinate out, 
then the $8$D worldvolume effective action for a D$7$-brane is obtained from the $10$D one (\ref{eq:10DSYM}) for a D$9$-brane.
From $10$D to $8$D, the gauge boson $A_M$ is decomposed into 
a complex scalar $\phi$, a $4$D spacetime vector $A_\mu$ ($\mu=0,1,2,3$) and an extra-dimensional 
vector $A_m$ ($m=4,5,6,7$), those are mode-expanded as
\begin{eqnarray}
\phi (x,x')&=&\sum_{i=-\infty}^\infty \phi^{(i)}(x)\Phi^{(i)}(x'),
\nonumber\\
A_\mu (x,x')&=& \sum_{i=-\infty}^\infty A_\mu^{(i)}(x)A^{(i)}(x'),
\nonumber\\
A_m (x,x')&=& \sum_{i=-\infty}^\infty \phi_m^{(i)}(x)A_m^{(i)}(x'),
\label{eq:decomposition}
\end{eqnarray}
where $x$ is the 4D Minkowski spacetime coordinate and $x'$ is the 
extra-dimensional coordinate on the 
D$7$-brane worldvolume, respectively. 

Next, in order to describe the mode expansion of the spinor field, 
we take the $10$D gamma matrices as shown in Ref.~\cite{Conlon:2008qi},
\begin{eqnarray}
\Gamma^\mu=\gamma^\mu \otimes \1 \otimes \1, \quad \Gamma^m=\gamma^5 \otimes \tilde{\gamma}^{m-3} \otimes \1, \quad \Gamma^p=\gamma^5 \otimes \tilde{\gamma^5} \otimes \tau^p,
\label{eq:10Dgamma}
\end{eqnarray}
where $p=8,9$. The $4$D Minkowski gamma matrices are expressed as
\begin{eqnarray}
\gamma^0 = \left(
\begin{array}{cc}
0&-\1 \\
\1 & 0
\end{array}
\right), \quad
\gamma^1 = \left(
\begin{array}{cc}
0&\sigma_x \\
\sigma_x & 0
\end{array}
\right), \quad
\gamma^2 = \left(
\begin{array}{cc}
0&\sigma_y \\
\sigma_y & 0
\end{array}
\right), \quad
\gamma^3 = \left(
\begin{array}{cc}
0&\sigma_z \\
\sigma_z & 0
\end{array}
\right),
\end{eqnarray}
and then $\gamma^5=i\gamma^0 \gamma^1 \gamma^2 \gamma^3$. On the other hand, the Euclidean gamma matrices for the internal 
coordinates of D$7$-brane are 
\begin{eqnarray}
\tilde{\gamma}^1 = \left(
\begin{array}{cc}
0&-i\1 \\
i\1 & 0
\end{array}
\right), \quad
\tilde{\gamma}^2 = \left(
\begin{array}{cc}
0&\sigma_z \\
\sigma_z & 0
\end{array}
\right), \quad
\tilde{\gamma}^3 = \left(
\begin{array}{cc}
0&\sigma_x \\
\sigma_x & 0
\end{array}
\right), \quad
\tilde{\gamma}^4 = \left(
\begin{array}{cc}
0&\sigma_y \\
\sigma_y & 0
\end{array}
\right),
\end{eqnarray}
and in this case $\tilde\gamma^5 =\tilde\gamma^1 \tilde\gamma^2 \tilde\gamma^3 \tilde\gamma^4$. 
The other directions described by the Pauli matrices, $\tau^8 = \sigma_x$ and $\tau^9 =\sigma_y$.

In this notation, the $10$D Majorana-Weyl spinor $\lambda$ can be decomposed as
\begin{eqnarray}
\lambda =(\lambda_1 +\lambda_4) \oplus (\lambda_2 +\lambda_3),
\end{eqnarray}
where
\begin{eqnarray}
\lambda_1 &=&\xi_1^+ (x) \psi_1^+ (x') \theta_1^+ (u),
\nonumber\\
\lambda_2 &=&\xi_2^+ (x) \psi_2^- (x') \theta_2^- (u),
\nonumber\\
\lambda_3 &=&\xi_3^- (x) \psi_3^- (x') \theta_3^+ (u),
\nonumber\\
\lambda_4 &=&\xi_4^- (x) \psi_4^+ (x') \theta_4^- (u), 
\label{eq:chirality}
\end{eqnarray}
$u$ is the transverse coordinate to the D$7$-brane, and the signs in the subscript of each factor express the chirality of spinor component the factor is contained with respect to the space(time) it feels. 
In the dimensional reduction from $10$D to $8$D as in the bosonic case, we assume the third factors in Eq.~(\ref{eq:chirality}) which depend on the transverse 
coordinate $u$ to the D$7$-brane are treated as constants in this paper.

The $10$D gauge coupling is contained in the second term of 
Eq.~(\ref{eq:10DSYM}) with the covariant derivative of Eq.~(\ref{eq:covariantD}) such as
\begin{eqnarray}
\int d^{10} x \{ \bar{\lambda} \Gamma^0 \Gamma^M [A_M,\lambda] \},
\label{eq:Yukawa10DSYM}
\end{eqnarray}
which leads to the $4$D Yukawa coupling for the transverse vector $A_p$ and/or the internal vector $A_m$, 
via the overlap integral of the extra dimensional factors shown in Eqs.~(\ref{eq:decomposition}) and~(\ref{eq:chirality}). 
For more details, see Ref.~\cite{Conlon:2008qi} and references therein.

Since the SYM gauge fields are represented by adjoint matrices of $U(N)$, we denote them as $\Phi_{{\frak a}{\frak b}}$ with the $U(N)$ index ${\frak a},{\frak b}=1,\cdots ,N$.
If the magnetic fluxes are turned on for the $U(1)$ subgroups, the original $U(N)$ breaks down to $U(N_1)\times U(N_2) \times \cdots \times U(N_n)$ with $N=N_1+N_2+\cdots +N_n$, 
and bifundamental representations of the product groups appear from the off-diagonal components of $\Phi_{{\frak a}{\frak b}}$. 
For example, when the flux $F_{{\frak a}{\frak b}} \propto {\rm diag}(m_1, \ldots,m_N)$ with $m_1=\cdots=m_{N_1} \equiv M_{(N_1)}$ and $m_{N_1+1}=\cdots=m_N \equiv M_{(N_2)}$ is 
turned on for the $U(1)$ subgroup of  $U(N_1+N_2)$, it induces the gauge symmetry breaking $U(N_1+N_2)\rightarrow U(N_1)\times U(N_2)$ 
and bifundamental fields $({\bf N}_1,\bar{\bf N}_2), (\bar{\bf N}_1, {\bf N}_2) $ are generated. 
In the Laplace and Dirac equations, the bifundamental field $({\bf N}_1,\bar{\bf N}_2)$ ($(\bar{\bf N}_1, {\bf N}_2)$) feels $M_{(N_1)}-M_{(N_2)}$ ($M_{(N_2)}-M_{(N_1)}$) units of flux. 
Then, as in the case of magnetized tori~\cite{Cremades:2004wa}, we expect the appearance of multiple chiral zero-modes, whose degree of degeneracy 
are determined by the number of flux themselves, are identified as the generation of matter fields. 

\section{Wavefunctions on the D$7$-branes with magnetic fluxes}
\label{sec:3}
In this paper, we restrict ourselves to the case $(n_1,n_2)=(1,1)$, 
for simplicity and the extension to the general integers $n_1$ and $n_2$ is straightforward. 
From Eq.~(\ref{eq:rpsi}) with $n_1=n_2=1$, the transverse position of the D$7$-brane expressed by 
coordinates $r$ and $\psi$ are dependent to its internal coordinates $\theta_i$ and $\phi_i$ as
\begin{eqnarray}
r^3&=&\left( \frac{c}{\sin\frac{\theta_1}{2}\sin\frac{\theta_2}{2}}\right)^2,
\nonumber\\
\psi &=&\phi_1 +\phi_2,
\label{eq:rpsi11}
\end{eqnarray}
in other words, it corresponds to the case that the D$7$-brane is spreading on the $z_1=c$ 
plane with the holomorphic coordinates shown in Eq.~(\ref{eq:z}). In this case the worldvolume coordinates of D$7$-brane are $z_3$ and $z_4$, those are expressed as
\begin{eqnarray}
z_3=c \cot \frac{\theta_1}{2} e^{i\phi_1},\quad z_4=c\cot \frac{\theta_2}{2} e^{i\phi_2}.
\label{eq:z3z4}
\end{eqnarray}
By substituting Eq.~(\ref{eq:rpsi11}) into Eq.~(\ref{eq:D7metric}), the induced metric is given by
\begin{eqnarray}
ds_{D7}^2 =\frac{4(|z_3|^2 +\frac{3}{2} c^2)}{9(|z_3|^2 +c^2)^2} |dz_3|^2 +\frac{4(|z_4|^2 +\frac{3}{2} c^2)}{9(|z_4|^2 +c^2)^2} |dz_4|^2 +\frac{4(\bar{z}_3z_4 dz_3 d\bar{z}_4+z_3\bar{z}_4 d\bar{z}_3 dz_4)}{9(|z_3|^2 +c^2)(|z_4|^2 +c^2)}. 
\label{eq:D7metricz3z4}
\end{eqnarray}
Note that from Eq.~(\ref{eq:rpsi11}),
\begin{eqnarray}
r^3=\frac{(|z_3|^2+c^2)(|z_4|^2+c^2)}{c^2},
\end{eqnarray}
the radial coordinate $r$ has a non-zero minimal value $r_{\rm min}=c^{2/3}$ in the limit $|z_3|,|z_4|\to 0$ as we mentioned before. 
However, $r$ diverges in the large radius limit $|z_3|,|z_4| \to \infty$, because the 
D$7$-brane extends to the region of global Calabi-Yau manifold, which is outside the boundary of near horizon limit. 
In what follows, we assume that the volume of D$7$-brane is almost 
determined by the four-cycle in the near horizon limit. 
Therefore, we define the finite value $z_{\rm max}$ as the 
boundary of near-horizon limit. 
In the limit $|z_3|,|z_4|\longrightarrow |z_{\rm max}|$, the radial coordinate $r$ reaches $r_{\rm max}=L^3=\displaystyle\frac{(|z_{\rm max}|^2+c^2)^2}{c^2}$.\footnote{Now it is assumed the maxima of $z_3$ and 
$z_4$ are the same as each other, for simplicity.} 
 
On the D$7$-brane, the K\"{a}hler form is given by pulling 
back from the original one in $10$D,
\begin{eqnarray}
J&=&ig_{i\bar{j}}dz^i \wedge d\bar{z}^{\bar{j}}
\nonumber\\
 &=&-\frac{1}{6}Q_1 \Omega_{11} -\frac{1}{6}Q_2\Omega_{22} -\frac{1}{9}\cot\frac{\theta_1}{2}\cot\frac{\theta_2}{2}\left(\Omega_{12} +\Omega_{21} \right),
\end{eqnarray}
where
\begin{eqnarray}
g_{i\bar{j}}=\frac{c^2}{3(|z_i|^2+c^2)^2}\delta_{i\bar{j}}+\frac{2}{9}\left( \frac{\bar{z_i}}{|z_i|^2+c^2} \right) \left( \frac{z_j}{|z_j|^2+c^2} \right),
\end{eqnarray}
and
\begin{eqnarray}
Q_i &=&\frac{3}{2} +\cot^2\frac{\theta_i}{2},
\nonumber\\
\Omega_{ij}&=&d\theta_i \wedge \sin \theta_j d\phi_j.
\end{eqnarray}
As discussed in the introduction, we introduce the 
two-form fluxes satisfying the Bianchi identity, i.e., 
\begin{eqnarray}
F=M_1Q_1 \Omega_{11} +M_2Q_2 \Omega_{22},
\label{eq:flux}
\end{eqnarray}
which are quantized on the basis of two-cycles 
$(\theta_i,\phi_i)$ as 
\begin{eqnarray}
\int_{(\theta_1,\phi_1)} F  &=&2\pi M_1{\cal V}_1=: 2\pi N_1,
\nonumber\\
\int_{(\theta_2,\phi_2)} F&=&2\pi M_2{\cal V}_2=: 2\pi N_2,
\label{eq:NormFlux}
\end{eqnarray}
with the volumes ${\cal V}_1$ and ${\cal V}_2$ of local two-cycles in the D$7$-brane given by,
\begin{eqnarray}
{\cal V}_1=\frac{L^2}{9}\int_{\theta_1^{\rm min}}^\pi d\theta_1 \sin\theta_1 Q_1,\,\,\,
{\cal V}_2=\frac{L^2}{9}\int_{\theta_2^{\rm min}}^\pi d\theta_2 \sin\theta_2 Q_2,
\end{eqnarray}
where $\theta_1^{\rm min}$ and $\theta_2^{\rm min}$ are determined by 
$|z_3|=|z_4|=|z_{\rm max}|$ with Eq.~(\ref{eq:z3z4}) as mentioned before. 
The Dirac quantization condition requires 
$M_i=N_i/{\cal V}_i$ $(N_i\in \mathbb{Z})$ with $i=1,2$ and the explicit 
forms of the field strengths are
\begin{eqnarray}
F_{z_3\bar{z}_3}=-2iM_1\frac{|z_3|^2 +\frac{3}{2}c^2}{(|z_3|^2+c^2)},\,\,\, F_{z_4\bar{z}_4}=-2iM_2\frac{|z_4|^2 +\frac{3}{2}c^2}{(|z_4|^2+c^2)},
\end{eqnarray}
those are supplied by
\begin{eqnarray}
A=iM_1\left( -\frac{c^2}{2z_3(|z_3|^2+c^2)}+\frac{1}{z_3}\ln (|z_3|^2+c^2) \right) dz_3\nonumber\\
-iM_1\left( -\frac{c^2}{2\bar{z}_3(|z_3|^2+c^2)}+\frac{1}{\bar{z}_3}\ln (|z_3|^2+c^2) \right) d\bar{z}_3\nonumber\\
+iM_2\left( -\frac{c^2}{2z_4(|z_4|^2+c^2)}+\frac{1}{z_4}\ln (|z_4|^2+c^2) \right) dz_4\nonumber\\
-iM_2\left( -\frac{c^2}{2\bar{z}_4(|z_4|^2+c^2)}+\frac{1}{\bar{z}_4}\ln (|z_4|^2+c^2) \right) d\bar{z}_4.
\end{eqnarray}

As for the two-form fluxes~(\ref{eq:flux}), there are two possibilities to satisfy the supersymmetric condition along the $D$-flat direction. 
First case is the supersymmetric fluxes $\int J \wedge F=0$ without any vacuum expectation values (VEVs) of matter fields, that is, $M_1=-M_2$ 
for the same size of local two-cycles ${\cal V}_1={\cal V}_2$ in the D$7$-brane. 
Second case is the non-supersymmetric fluxes $\int J\wedge F\neq 0$, i.e., $M_1\neq -M_2$ for ${\cal V}_1={\cal V}_2$, which are expected (assumed) 
to be canceled by some non-vanishing VEVs of charged scalar fields under the fluxed $U(1)$ symmetry. 
In the following, we consider these two supersymmetric cases. 

Along the line of Ref.~\cite{Conlon:2008qi}, we solve the equations of motion for scalars and fermions in 
the next subsections. 
Before going to the detail of them, in the following, we comment on a shift of the number of magnetic fluxes, referred to as twisting, that encodes possible curvature effects in transverse directions to D-branes.
The equation of motion for the scalar mode obeys the Laplace equation,
\begin{eqnarray}
-\tilde{D}_m \tilde{D}^m \Phi =m^2 \Phi,
\end{eqnarray}
where $\tilde{D}_m$ is the covariant derivative, $\tilde{D}_m \Phi =\nabla_m \Phi -i[A_m, \Phi]$.
One of scalar modes for the extra-dimensional space comes from the $4$D gauge fields (vector degree of freedom in Minkowski spacetime) of the D$7$-branes
and it has a value in the tangent bundle of the D$7$-brane. 
The other is coming from the transverse scalar mode (position moduli) which has a value in the normal bundle.
The dimensional reduction of the maximal SYM theory
from $10$D to $8$D preserving the supersymmetry decomposes its field contents to an $8$D gauge boson, a gaugino, a complex scalar and its superpartner fermion.
In flat space, these transform under the representation for $SO(3,1)\times SO(4)\times U(1)_R$ and
fermions have $\pm 1/2$ charge under $U(1)_R$.
The new central $U(1)$ charge embedded in $SO(4)$ in general gives rise to the twisted SYM theory on 
the D$7$-branes~\cite{Beasley:2008dc}, in order to allow the four scalar superchages in $4$D Minkowski spacetime. In our case, these isometies are determined by the way of locating D$7$-branes in the conifold shown in Eq.~(\ref{eq:rpsi}), however, those could be further modified if it is embedded into a global CY space as discussed in Sec.~\ref{subsec:global}.
When a nontrivial curvature of the normal bundle exists, there appears an additional effect in the covariant derivatives and the equations of motion are modified.
The additional terms are proportional to the K\"{a}hler form which consequently shift the effective number of magnetic fluxes felt by each field. Such effects are called twisting. 

On the other hand, the vector modes in the extra-dimensional space, called the internal vector modes (Wilson-line moduli), correspond to the scalar degrees of freedom in $4$D Minkowski spacetime. 
Their equations of motion are extracted as
\begin{eqnarray}
\tilde{D}_m \tilde{D}^m \Phi_n +2iF^m_{\ n} \Phi_m -[\nabla^m, \nabla_n] \Phi_n =-m^2 \Phi_m,
\label{eq:vector}
\end{eqnarray}
from which we find these modes do not receive the twisting due to the structure of the tangent bundle. 

Finally, the equation of motion for the fermion obey the Dirac equation,
\begin{eqnarray}
\Gamma^m \tilde{D}_m \Psi =0,
\end{eqnarray}
where the covariant derivative $\tilde{D}_m$ includes the spin connection term. 
The fermionic modes are also affected by the twist, i.e., the shift of magnetic flux. 

In this paper, we start from $10$D SYM action~(\ref{eq:10DSYM}) and dimensionally reduce it to $8$D one on the D$7$-brane which should preserve a supersymmetry ensured by the kappa-symmetry.
We do not identify the explicit forms of twisting for fields on D$7$-branes~(\ref{eq:decomposition}) and~(\ref{eq:chirality}) in the $AdS_5 \times T^{1,1}$ local spacetime around the tip of conifold, because they could be further modified when we embed it into a global CY, and treat the shifts of magnetic fluxes as parameters in our local model. 
Then we count the numbers of zero-modes which localize toward the tip by assuming the minimal twist in each sector\footnote{In other words, we implicitly consider a global CY space which provides the minimal twist in the conifold region.} and study their wavefunctions in the conifold region. Because our analysis is valid only in such a region inside the boundary of near horizon limit, in the following, we just abandon the other zero-modes which localize toward the opposite side to the tip of conifold, and focus on analyses for those localize toward the tip. Note that, when the conifold is embedded into a global CY space, the fields localize against the tip should be in general affected by the detailed structure of CY. Furthermore, it might be also possible that the embedding yields new zero-modes outside the boundary of near horizon limit. Since D7-branes are located in a kappa-symmetric way in the local conifold region, 
the structure of $4$D ${\cal N}=1$ supersymmetry would be also captured by focusing on the massless bosons and fermions as analyzed later.

\subsection{Bosons}
\label{subsec:bosons}
\subsubsection{Transverse scalar modes}
The scalar spectrum for the extra-dimensional part on the 
D$7$-brane, transverse scalar mode and Minkowski vector mode are derived by solving the Laplace equation,
\begin{eqnarray}
-g^{mn}D_m D_n \Phi &=&-2\sum_{i,j=3,4}g^{z_i\bar{z}_j}D_{z_i}D_{\bar{z}_j} \Phi -\sum_{i,j=3,4}g^{\bar{z}_iz_j}[D_{\bar{z}_i}, D_{z_j}]\Phi =m^2 \Phi,
\label{eq:LaplaceEq}
\end{eqnarray}
where $D_{z_i}=\partial_{z_i}-iA_{z_i}$ is the covariant derivative, $D_{\bar{z_i}}$ is its Hermitian conjugate, and the eigenvalue $m^2$ corresponds to a mass-squared of the eigenmode. 
For the supersymmetric fluxes, $M_1=-M_2$, the term in Eq.~(\ref{eq:LaplaceEq}) with the commutator of $D_z$ and $D_{\bar{z}}$ satisfies
\begin{eqnarray}
\sum_{i,j=3,4}g^{\bar{z}_iz_j}[D_{\bar{z}_i}, D_{z_j}]\Phi =-i\sum_{i,j=3,4}g^{\bar{z}_iz_j}F_{z_j\bar{z}_i}\Phi =0.
\label{eq:fluxcond}
\end{eqnarray}
On the other hand, for the non-supersymmetric cases, $M_1 \ne -M_2$, the term~(\ref{eq:fluxcond}) is nonvanishing but it is canceled by the non-zero VEVs of some scalar fields at the Lagrangian level, 
then subtracted from the mode equation~(\ref{eq:LaplaceEq}). 
Therefore, in both cases, the equation $D_{z_i}\Phi=0$ or $D_{\bar{z}_i}\Phi=0$ ($i=3,4$) determines wavefunction for a massless zero-mode. Our goal is obtaining the particular 
solution of them. 
(In local $\mathbb{P}^2$ model~\cite{Conlon:2008qi} with supersymmetric fluxes, Eq.~(\ref{eq:fluxcond}) becomes non-zero and there is only massive scalar mode after twisting.) 
By solving the parts of zero-mode equations $D_{z_3}\Phi_1 =0$, $D_{\bar{z}_3}\Phi_2 =0$, $D_{z_4}\Phi_3 =0$ and $D_{\bar{z}_4}\Phi_4 =0$, we obtain
\begin{eqnarray}
\Phi_1 &=&f(\bar{z}_3)e^{M_1\left( -\ln c^2 \ln z_3 + {\rm Li}_2(-\frac{|z_3|^2}{c^2}) \right)}\left( \frac{z_3}{|z_3|^2+c^2}\right)^{\frac{M_1}{2}},
\nonumber\\
\Phi_2 &=&g(z_3)e^{-M_1\left( -\ln c^2 \ln \bar{z}_3 + {\rm Li}_2(-\frac{|z_3|^2}{c^2}) \right)}\left( \frac{\bar{z}_3}{|z_3|^2+c^2}\right)^{-\frac{M_1}{2}},
\nonumber\\
\Phi_3 &=&h(\bar{z}_4)e^{M_2\left( -\ln c^2 \ln z_4 + {\rm Li}_2(-\frac{|z_4|^2}{c^2}) \right)}\left( \frac{z_4}{|z_4|^2+c^2}\right)^{\frac{M_2}{2}},
\nonumber\\
\Phi_4 &=&k(z_4)e^{-M_2\left( -\ln c^2 \ln \bar{z}_4 + {\rm Li}_2(-\frac{|z_4|^2}{c^2}) \right)}\left( \frac{\bar{z}_4}{|z_4|^2+c^2}\right)^{-\frac{M_2}{2}},
\label{eq:bosonwave}
\end{eqnarray}
where $g(z_3)$ and $k(z_4)$ ($f(\bar{z}_3)$ and $h(\bar{z}_4)$) are holomorphic (anti-holomorphic) functions constrained by the normalization of wavefunctions. In the exponents, ${\rm Li}_2 (z)$ represents the dilogarithm (Spence) function defined by 
\begin{eqnarray}
{\rm Li}_2(z)=-\int_{0}^1 \frac{\ln (1-zt)}{t} dt,
\end{eqnarray}
where $z$ is a complex number, which has a cut along the real axis of $z$ from $1$ to $\infty$. This sort of integral is also employed in Feynman parameter integrals for a one-loop amplitude~\cite{'tHooft:1978xw}. 
Using two of four functions in Eq.~(\ref{eq:bosonwave}), we can construct two ansatzes, 
$\Phi =\Phi_1 \Phi_3$ or $\Phi_2 \Phi_4$, for the scalar wavefunction. 

When we impose the periodic boundary condition $\Phi (\theta_i, \phi_i +2\pi)=\Phi (\theta_i, \phi_i)$,  wavefunctions are proportional to the integer power of $z_i$ and/or $\bar{z}_i$. Therefore, the arbitrary functions in Eq.~(\ref{eq:bosonwave}) are constrained as 
\begin{eqnarray}
f(\bar{z}_3)&=&\sum_{p\in {\mathbb Z}}A_p^f \bar{z}_3^{p+M_1(-\ln c^2 +1/2)},\nonumber\\
g(z_3)&=&\sum_{p'\in {\mathbb Z}}A_{p'}^g z_3^{p'-M_1(-\ln c^2 +1/2)},\nonumber\\
h(\bar{z}_4)&=&\sum_{q\in {\mathbb Z}}A_q^h \bar{z}_4^{q+M_2(-\ln c^2 +1/2)},\nonumber\\
k(z_4)&=&\sum_{q'\in {\mathbb Z}}A_{q'}^k z_4^{q'-M_2(-\ln c^2 +1/2)},
\label{eq:Fourier}
\end{eqnarray}
where the integers $p$, $p'$, $q$, $q'$ and real constants $A_{p,p',q,q'}^{f,g,h,k}$ are determined by the normalization condition discussed later. 
By substituting Eq.~(\ref{eq:Fourier}) to Eq.~(\ref{eq:bosonwave}), each mode of the bosonic zero-mode wavefunctions $\Phi_{13,pq}\equiv \Phi_{1,p}\Phi_{3,q}$ and $\Phi_{24,p'q'}\equiv \Phi_{2,p'}\Phi_{4,q'}$ is rewritten to be
\begin{eqnarray}
\Phi_{1,p} &=&A_p^f e^{M_1 {\rm Li}_2(-\frac{|z_3|^2}{c^2})}\left( |z_3|^2+c^2\right)^{-\frac{M_1}{2}}|z_3|^{2M_1\left( -\ln c^2 +1/2\right)}\bar{z}_3^p,
\nonumber\\
\Phi_{2,p'} &=&A_{p'}^g e^{-M_1 {\rm Li}_2(-\frac{|z_3|^2}{c^2})}\left( |z_3|^2+c^2\right)^{\frac{M_1}{2}}|z_3|^{-2M_1\left( -\ln c^2 +1/2\right)}z_3^{p'},
\nonumber\\
\Phi_{3,q} &=&A_q^h e^{M_2 {\rm Li}_2(-\frac{|z_4|^2}{c^2})}\left( |z_4|^2+c^2\right)^{-\frac{M_2}{2}}|z_4|^{2M_2\left( -\ln c^2 +1/2\right)}\bar{z}_4^q,
\nonumber\\
\Phi_{4,q'} &=&A_{q'}^k e^{-M_2 {\rm Li}_2(-\frac{|z_4|^2}{c^2})}\left( |z_4|^2+c^2\right)^{\frac{M_2}{2}}|z_4|^{-2M_2\left( -\ln c^2 +1/2\right)}z_4^{q'}.
\label{eq:bw}
\end{eqnarray}
Since the different modes of wavefunctions $\Phi_{13,pq}$ ($\Phi_{24,p'q'}$) are orthogonal to each other, i.e., 
$\int dz_3^2dz_4^2 \sqrt{g}\Phi_{13,p_1q_1}^\dagger\Phi_{13,p_2q_2}=0$ for $p_1 \neq p_2$ or $q_1 \neq q_2$ (similarly for $\Phi_{24,p'q'}$) due to the periodicity, they are independent solutions for the Laplace equation~(\ref{eq:LaplaceEq}) labelled by the integers $p$, $p'$, $q$ and $q'$.
We will see below that the normalization and validity conditions of wavefunctions in the conifold region restrict the allowed integers for $p$, $p'$, $q$ and $q'$, and the number of possible combinations of these integers corresponds to the degeneracy of zero-modes, which is identified as the number of bosonic generations.
The different types of normalization conditions with and without fluxes are categorized into the following three cases.

\medskip

\noindent
$\bullet \quad M_1=-M_2 \equiv M \ne 0$

\medskip
First, we show the normalization of $\Phi_{13,pq}$ with the supersymmetric fluxes $M_1=-M_2 \equiv M \ne 0$. The 
normalization condition is defined by
\begin{eqnarray}
\int d^2 z_3 d^2 z_4 \sqrt{g}\Phi_{13,p_iq_i}^\dagger \Phi_{13,p_jq_j}=\delta_{p_ip_j}\delta_{q_iq_j},
\end{eqnarray}
which is explicitly calculated as
\begin{eqnarray}
1=
\frac{16\pi^2 c^2}{27}(A_p^f)^2(A_q^h)^2\int dr_3 dr_4
e^{2M{\rm Li}_2(-\frac{r_3^2}{c^2})}e^{-2M{\rm Li}_2(-\frac{r_4^2}{c^2})}
\frac{2r_3^2+2r_4^2+3c^2}{(r_3^2+c^2)^{M+2}(r_4^2+c^2)^{-M+2}}\nonumber\\
\times r_3^{2p+1+4M(-\ln c^2 +1/2)} r_4^{2q+1-4M(-\ln c^2 +1/2)}.
\label{eq:MbosN}
\end{eqnarray}
From the asymptotic expansion of the dilogarithm function in the limit $r\gg 1$, 
\begin{eqnarray}
{\rm Li}_2\left(-\frac{r^2}{c^2}\right)\simeq -\frac{\pi^2}{6}-\frac{1}{2}\left(\ln \left(\frac{r^2}{c^2}\right)\right)^2 +{\cal O}\left(\frac{c^2}{r^2}\right),
\label{eq:dialog}
\end{eqnarray}
the factor $e^{-2M{\rm Li}_2(-r_4^2/c^2)}$ ($e^{2M{\rm Li}_2(-r_3^2/c^2)}$) in the integrand of the Eq.~(\ref{eq:MbosN}) diverges in the limit $r_4\gg c$ ($r_3\gg c$) if the flux $M$ is chosen as a positive (negative) value. 
Therefore the bosonic wavefunction with the supersymmetric flux is always non-normalizable. %although it is normalizable for the finite ranges 
%$z_3,z_4\in [0,z_{max}]$. 
A similar analysis can be performed for the other wavefunction $\Phi_{24,p'q'}$, and the result is the same as $\Phi_{13,pq}$. 

\medskip

\noindent
$\bullet \quad M_1=M_2 \equiv M \ne 0$

\medskip

In contrast to the supersymmetric fluxes, here we consider the non-supersymmetric flux $M_1=M_2 \equiv M \ne 0$, that is, the different sign of $M_2$ in Eq.~(\ref{eq:flux}) from the supersymmetric one, such as
\begin{eqnarray}
F=MQ_1 \Omega_{11} +MQ_2 \Omega_{22},
\label{eq:FI}
\end{eqnarray}
which leads to the non-zero Fayet-Iliopoulos (FI) term. 
Such a non-vanishing FI term can be canceled by some VEVs of charged scalar fields under the fluxed $U(1)$ symmetry, and we assume such a situation in this paper. 
Because of the sign flip $-M_2 \to M_2$ from the previous supersymmetric case $M_1=-M_2 \ne 0$, the bosonic wavefunction becomes normalizable as shown below. 

At the beginning, we discuss about the upper bound of $p$ and $q$ to have normalizable solutions, and next we show the lower bound of them. 
First of all, our analysis relies on the approximation that the tail of wavefunction outside the boundary of near horizon limit does not cause sizable effects. 
Thus, the wavefunction have to be localized around the tip of cone, otherwise it cannot be controlled. 
The bosonic wavefunction in this case is asymptotically given in the limit $R=r_3=r_4\gg c$ as 
\begin{eqnarray}
\hat{\Phi}_{13,pq}\equiv 4\pi R(\sqrt{g})^{1/2}|\Phi_{13,pq}|&\propto& R^{p+q-4M\ln c^2-2}e^{-M(\ln (R^2/c^2))^2}\nonumber\\
 &=& R^{p+q-2M\ln (c^2R^2)-2}c^{2M\ln (R^2/c^2)},
\end{eqnarray}
by employing Eq.~(\ref{eq:dialog}) and then the extremal condition of $\hat{\Phi}_{13,pq}$ is achieved by the following relation,
\begin{eqnarray}
p+q-2-8M\ln R_{\ast}=0,
\label{eq:bsloc}
\end{eqnarray}
where $R_\ast$ represents the extremal point, around which the wavefunction localizes. 
In order to obtain the localized wavefunction around the tip of cone, we require $R_\ast \ll \sqrt{cL^{3/2}-c^2}$, \footnote{$r_{\rm max}=L^3 \gg \frac{(R_{\ast}^2+c^2)^2}{c^2}$.} i.e., 
\begin{eqnarray}
p+q <8M\ln \sqrt{cL^{3/2}-c^2} +2,
\label{eq:gen}
\end{eqnarray}
which shows a validity condition of the bosonic wavefunction in the local conifold region. 

With the typical parameters such as the string coupling $g_s$ and the number of D$3$-branes $N_c$ given by 
\begin{eqnarray}
g_s=0.1,\qquad N_c=100,
\label{eq:para1}
\end{eqnarray}
ensuring that the backreaction of D$7$-brane is negligible, the horizon scale $L$ is taken as 
\begin{eqnarray}
7.3 \leq L \leq 146,
\end{eqnarray}
in the unit of $M_{\rm Pl}=1$, within the range of the string scale $M_s=1/(2\pi \sqrt{\alpha^\prime})$, $10^{16}\,{\rm GeV}\leq M_s \leq 2\times 10^{17}\,{\rm GeV}$. 
For example, when we choose the string scale as $M_s\simeq 1.08\times 10^{17}\,{\rm GeV}$,  the validity condition~(\ref{eq:gen}) is evaluated as 
\begin{eqnarray}
p+q <2+0.8 N\,\,\,(L\simeq 13.5),
\label{eq:BosonConv}
\end{eqnarray}
with $c=1$ and quantized fluxes $N=N_1=N_2$. The certain choice of $L$ determines the value of $\theta_{\rm min}$ and quantized fluxes $N$ can be calculated by Eq.~(\ref{eq:NormFlux}). 

On the other hand, the lower bounds of $p$ and $q$ are determined by the convergence condition of the normalization factors derived from Eq.~(\ref{eq:MbosN}) with the flipped sign of $M_2$. Around the origin of $r_3$ and $r_4$, the integrand of Eq.~(\ref{eq:MbosN}) is asymptotically given by
\begin{eqnarray}
c^{-4M-8} (2r_3^2 +2r_4^2+3c^2) r_3^{2p+1+4M(-\ln c^2 +1/2)}r_4^{2q+1+4M(-\ln c^2 +1/2)},
\end{eqnarray}
and the normalizable bosonic wavefunction requires
\begin{eqnarray}
p>-1-2M(-\ln c^2 +\frac{1}{2}), \quad q>-1-2M(-\ln c^2 +\frac{1}{2}).
\label{eq:gen2}
\end{eqnarray}
For the string scale $M_s \simeq 1.08 \times 10^{17}$ GeV, the lower bounds~(\ref{eq:gen2}) are expressed as 
\begin{eqnarray}
p >-1-0.05 N, \quad q >-1-0.05 N\,\,\,(L\simeq 13.5).
\label{eq:BosonConv2}
\end{eqnarray}
The existence of upper and lower bounds for $p$ and $q$ implies that there is finite number of degenerate zero-modes, those are distinguished by different combinations of $p$ and $q$  from each other allowed by the condition~(\ref{eq:BosonConv}) and~(\ref{eq:BosonConv2}) with the fixed number of flux $N$.

\begin{figure}
\centering \leavevmode
\includegraphics[width=90mm]{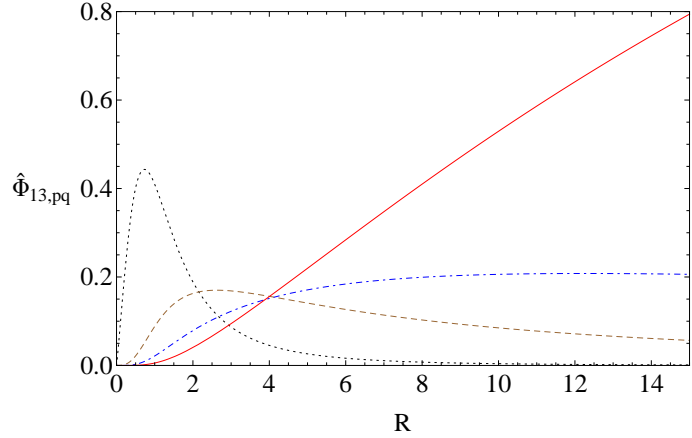}
\caption{The typical bosonic wavefunction $\hat{\Phi}_{13,pq}$ for the non-supersymmetric fluxes $N_1=N_2=1$, on the $R=r_3=r_4$ hypersurface. 
The black dotted ($p=q=0$) and the brown dashed ($p=q=1$) curves are localized at the point inside the boundary of near horizon limit, $R\lesssim 7$ in contrast to the blue dot-dashed ($p=1,q=2$) and red solid curves ($p=q=2$), those are localized outside it.}
\label{fig:bs_FI}
\end{figure}

To confirm the above statements, we exhibit the bosonic wavefunction $\hat{\Phi}_{13,pq}$ on the $R=r_3=r_4$ hypersurface as shown in Fig.~\ref{fig:bs_FI}, where the parameters are chosen as 
\begin{eqnarray}
c=1,\,\,\, N=1,\,\,\, L\simeq 13.5,\,\,\, M_s=1.08\times 10^{17}\,{\rm GeV},
\label{eq:para2}
\end{eqnarray}
in the unit $M_{\rm Pl}=1$ with the typical parameters~(\ref{eq:para1}) and normalization factors are computed numerically. 
In Fig.~\ref{fig:bs_FI}, the wavefunctions drawn by the blue dot-dashed and the red-solid curves are those out of control, because each one localizes outside the boundary of near horizon limit, $R_\ast>\sqrt{cL^{3/2}-c^2}\simeq 7$. Such an observation is consistent with the above argument leading to Eq.~(\ref{eq:bsloc}). 
Therefore, as shown in Fig.~\ref{fig:bs_FI}, in this case there are fifteen independent zero-mode solutions, such as $(p,q)=(\pm 1,\pm1)$, $(\pm1,0)$, $(0,\pm1)$, $(0,0)$, $(-1,2)$, $(-1,3)$, $(0,2)$, $(2,-1)$, $(2,0)$ and $(3,-1)$, those yield fifteen generations of massless scalars. 
A similar analysis for the other wavefunction $\Phi_{24,p'q'}$ but with the replacement $M_2 \to -M_2$ can be performed, and the result is the same as $\Phi_{13,pq}$. 

Finally we focus the effect of twisting.
If the cycle wrapped D$7$-brane has nontrivial normal bundle, it causes the Laplace equation and the Dirac equation twist. 
In twisted equation of motion, magnetic fluxes that the bi-fundamental matter fields feel shift as $N\rightarrow N+1$.\footnote{We assume that the magnetic flux is shifted as $N\rightarrow N+1$, but actually the explicit form of shifting is determined by the topology of global CY. For example, $N\rightarrow N-3$ for bosons and $N\rightarrow N\pm 3/2$ for fermions in $\mathbb{P}^2$~\cite{Conlon:2008qi}. } Accordingly, the number of zero-modes is changed. This effect is also applied for the following zero-flux case.
\medskip

\noindent
$\bullet \quad M_1=M_2=0$

\medskip

The bosonic wavefunctions without the flux, i.e., $M_1=M_2=0$ are normalized by
\begin{eqnarray}
\int d^2 z_3 d^2 z_4 \sqrt{g}\Phi_{13,pq}^\dagger \Phi_{13,pq}&=&
\frac{16\pi^2 c^2}{27}(A_p^f)^2(A_q^h)^2\int dr_3 dr_4 \frac{2r_3^2+2r_4^2+3c^2}{(r_3^2+c^2)^2(r_4^2+c^2)^2} r_3^{2p+1} r_4^{2q+1}
\nonumber\\
&=&(A_p^f)^2(A_q^h)^2\frac{4\pi^4 c^{2(p+q)}(-pq-2p-2q)}{27\sin(p\pi)\sin(q\pi)},
\label{eq:0bosN}
\end{eqnarray}
where $z_i =r_i \exp [i\theta_i]$ for $i=3,4$. Then we obtain the normalization factors,
\begin{eqnarray}
A_p^{f} A_q^{h}= \sqrt{\frac{27\sin(p\pi)\sin(q\pi)}{4\pi^4 c^{2(p+q)}(-pq-2p-2q)}}.
\end{eqnarray}
Note that the integrals in Eq.~(\ref{eq:0bosN}) are performed over the radius coordinates $r_3$ and $r_4$ from $0$ to infinity. 
Then the following equality can be utilized in the evaluation of the integral in Eq.~(\ref{eq:0bosN}): 
\begin{eqnarray}
\int dr\frac{r^{2K+1}}{(r^2+c^2)^{L+1}}&=&c^{2K-2L}\int_0^{\frac{\pi}{2}} \sin^{2K+1}\theta \cos^{2L-2K-1}\theta
\nonumber\\ 
&=&c^{2K-2L}\frac{\Gamma (K+1)\Gamma( L-K)}{2\Gamma( L+1)}.
\end{eqnarray}
The normalization condition requires the allowed range of $p$ and $q$ as $-1<p<0$ and $-1<q<0$, otherwise the normalization factor diverges.
As a result, these range do not include integers $p$ and $q$, and 
there are no normalizable modes in this case.
A similar analysis leads to the same result also in the case of $\Phi_{24,p'q'}$. 

Actually, we are interested in the wavefunctions inside the boundary of near horizon limit.
The final expression in Eq.~(\ref{eq:0bosN}) can be considered as an approximated estimation, because in the current case the wavefunction $\Phi$ is localized in the radial coordinates $r_3$ and $r_4$, towards the opposite side to the near horizon, around the tip of cone.
As we discussed in the case of non-supersymmetric flux, 
with employing validity condition~(\ref{eq:gen}) and the convergence condition of the normalization factors~(\ref{eq:gen2}), there are three independent zero-mode solutions without flux, such as $(p,q)=(0,0)$, $(0,1)$ and $(1,0)$.
\subsubsection{Minkowski vector modes}
The extra-dimensional wavefunctions of the $4$D gauge fields on D$7$-branes (vector degrees of freedom in $4$D Minkowski spacetime) also obey Eq.~(\ref{eq:bw}), the same one as transverse scalar modes. Because the vector field in Minkowski spacetime is real valued, the wavefunction of each component should take a real number, that is, only $(p,q)=(0,0)$ is allowed. As a result, there is a single zero-mode solution for the Minkowski vector mode. Note that the wavefunction with $(p,q)=(0,0)$ becomes a constant, that implies the $4$D gauge field does not localize in the world-volume of D$7$-branes where it originates.
\subsubsection{Internal vector modes}
Next, let us focus on the wavefunction for the internal vector modes given by the Eq.~(\ref{eq:vector}). 
We impose the gauge fixing condition,
\begin{eqnarray}
\tilde{D}_m \Phi^m = g^{z_3 \bar{z}_3}\left( D_{z_3}\Phi_{\bar{z}_3} +D_{\bar{z}_3}\Phi_{z_3}\right) +g^{z_3 \bar{z}_4}\left( D_{z_3}\Phi_{\bar{z}_4} +D_{\bar{z}_4}\Phi_{z_3}\right) \nonumber\\
+g^{z_4 \bar{z}_3}\left( D_{z_4}\Phi_{\bar{z}_3} +D_{\bar{z}_3}\Phi_{z_4}\right) +g^{z_4 \bar{z}_4}\left( D_{z_4}\Phi_{\bar{z}_4} +D_{\bar{z}_4}\Phi_{z_4}\right), 
\end{eqnarray}
and especially concentrate on the solutions that both $\Phi_{z_3}=\Phi_{z_4}=0$ and $D_{z_3}\Phi_{\bar{z}_3}=D_{z_4}\Phi_{\bar{z}_3}=D_{z_3}\Phi_{\bar{z}_4}=D_{z_4}\Phi_{\bar{z}_4}=0$ hold 
(also flipping with $z_3\leftrightarrow \bar{z}_3$ and $z_4\leftrightarrow \bar{z}_4$). 
For the case of supersymmetric fluxes, the equation of motion~(\ref{eq:vector}) implies that the vector modes 
which satisfying the above conditions are just the zero-mode solutions as shown in Eq.~(\ref{eq:bosonwave}),
\begin{eqnarray}
\Phi_{z_3}&=&A^{(z_3)}(z_3,z_4)e^{-M_1\left( -\ln c^2 \ln \bar{z}_3 + {\rm Li}_2(-\frac{|z_3|^2}{c^2}) \right)}\left( \frac{\bar{z}_3}{|z_3|^2+c^2}\right)^{-\frac{M_1}{2}}
\nonumber\\
&\times &
e^{-M_2\left( -\ln c^2 \ln \bar{z}_4 + {\rm Li}_2(-\frac{|z_4|^2}{c^2}) \right)}\left( \frac{\bar{z}_4}{|z_4|^2+c^2}\right)^{-\frac{M_2}{2}}
 \quad (M_1, M_2\leq 0),
\nonumber\\
\Phi_{z_4}&=&A^{(z_4)}(z_3,z_4)e^{-M_1\left( -\ln c^2 \ln \bar{z}_3 + {\rm Li}_2(-\frac{|z_3|^2}{c^2}) \right)}\left( \frac{\bar{z}_3}{|z_3|^2+c^2}\right)^{-\frac{M_1}{2}}
\nonumber\\
&\times &
e^{-M_2\left( -\ln c^2 \ln \bar{z}_4 + {\rm Li}_2(-\frac{|z_4|^2}{c^2}) \right)}\left( \frac{\bar{z}_4}{|z_4|^2+c^2}\right)^{-\frac{M_2}{2}}
 \quad (M_1, M_2\leq 0),
\nonumber\\
\Phi_{\bar{z}_3}&=&A^{(\bar{z}_3)}(\bar{z}_3,\bar{z}_4) e^{M_1\left( -\ln c^2 \ln z_3 + {\rm Li}_2(-\frac{|z_3|^2}{c^2}) \right)}\left( \frac{z_3}{|z_3|^2+c^2}\right)^{\frac{M_1}{2}} 
\nonumber\\
& \times &
e^{M_2\left( -\ln c^2 \ln z_4 + {\rm Li}_2(-\frac{|z_4|^2}{c^2}) \right)}\left( \frac{z_4}{|z_4|^2+c^2}\right)^{\frac{M_2}{2}}
 \quad (M_1, M_2\geq 0),
\nonumber\\
\Phi_{\bar{z}_4}&=& A^{(\bar{z}_4)}(\bar{z}_3,\bar{z}_4) e^{M_1\left( -\ln c^2 \ln z_3 + {\rm Li}_2(-\frac{|z_3|^2}{c^2}) \right)}\left( \frac{z_3}{|z_3|^2+c^2}\right)^{\frac{M_1}{2}} 
\nonumber\\
& \times &
e^{M_2\left( -\ln c^2 \ln z_4 + {\rm Li}_2(-\frac{|z_4|^2}{c^2}) \right)}\left( \frac{z_4}{|z_4|^2+c^2}\right)^{\frac{M_2}{2}}
 \quad (M_1, M_2\geq 0),
\end{eqnarray}
where 
\begin{eqnarray}
A^{(z_3)}(z_3,z_4) &=& \sum_{p', q'\in {\mathbb Z}}A_{p',q'}^{(z_3)} z_3^{p'-M_1(-\ln c^2 +1/2)} z_4^{q'-M_2(-\ln c^2 +1/2)},
\nonumber\\
A^{(z_4)}(z_3,z_4) &=& \sum_{p', q'\in {\mathbb Z}}A_{p',q'}^{(z_4)} z_3^{p'-M_1(-\ln c^2 +1/2)} z_4^{q'-M_2(-\ln c^2 +1/2)},
\nonumber\\
A^{(\bar{z}_3)}(\bar{z}_3,\bar{z}_4) &=& \sum_{p', q'\in {\mathbb Z}}A_{p',q'}^{(\bar{z}_3)} \bar{z}_3^{p+M_1(-\ln c^2 +1/2)} \bar{z}_4^{q+M_2(-\ln c^2 +1/2)},
\nonumber\\
A^{(\bar{z}_4)}(\bar{z}_3,\bar{z}_4) &=& \sum_{p', q'\in {\mathbb Z}}A_{p',q'}^{(\bar{z}_4)} \bar{z}_3^{p+M_1(-\ln c^2 +1/2)} \bar{z}_4^{q+M_2(-\ln c^2 +1/2)}.
\end{eqnarray}
The integers $p'$, $q'$ and real constants $A_{p',q'}^{(z_3,z_4,\bar{z}_3,\bar{z}_4)}$ are determined by the normalization condition that 
\begin{eqnarray}
\int d^2 z_3 d^2 z_4 \sqrt{g} g^{i\bar{j}} \Phi_{i}^{\frak{ab}} \Phi_{\bar{j}}^{\frak{ba}} =1.
\label{eq:vectornorm}
\end{eqnarray}
In the case of non-supersymmetric flux $M_1 = M_2 \equiv M$, a relation $\Phi_{z_3}=\Phi_{z_4}$ is expected with $A^{(z_3)}=A^{(z_4)}\equiv A^{(z)}$ and similarly $\Phi_{\bar{z}_3}=\Phi_{\bar{z}_4}$ with $A^{(\bar{z}_3)}=A^{(\bar{z}_4)}\equiv A^{(\bar{z})}$ 
those are suggested by the symmetry under the exchange of $z_3$ and $z_4$. 
There is a constant solution for a normalizable zero-mode wavefunction with the vanishing flux, and for $M_1, M_2 \geq 0$, the left-hand side of Eq.~(\ref{eq:vectornorm}) is deformed to 
\begin{align}
&\int d^2 z_3 d^2 z_4\left( g_{z_3 \bar{z}_3} +g_{z_4 \bar{z}_4} - g_{z_3 \bar{z}_4} - g_{z_4 \bar{z}_3} \right) \Phi_{\bar{z}_3}^2 \nonumber\\
&=\int d^2 z_3 d^2 z_4 \left[ {\cal R}_1 (|z_3|,|z_4|) \bar{z}_3^{2p} \bar{z}_4^{2q} +{\cal R}_2 (|z_3|,|z_4|) (z_3 \bar{z}_4 +\bar{z}_3 z_4)\bar{z}_3^{2p} \bar{z}_4^{2q} \right],
\end{align}
where ${\cal R}_1$ and ${\cal R}_2$ are real-valued functions which take non-vanishing values only if $p=q=0$.
The wavefunction of the vector mode is asymptotically given in the limit $R=r_3=r_4\gg c$ as 
\begin{eqnarray}
\hat{\Phi}_{\bar{z}_3,pq}\equiv 4\pi R\left( g_{z_3 \bar{z}_3} +g_{z_4 \bar{z}_4} - g_{z_3 \bar{z}_4} - g_{z_4 \bar{z}_3} \right)^{1/2}\Phi_{\bar{z}_3,pq}&\propto& R^{-4M\ln c^2}e^{-M(\ln (R^2/c^2))^2}\nonumber\\
 &=& R^{-2M\ln (c^2R^2)}c^{2M\ln (R^2/c^2)},
\end{eqnarray}
and then the extremal condition of $\hat{\Phi}_{z_3,pq}$ is achieved by the following relation,
\begin{eqnarray}
 R_{\ast}=1,
\label{eq:vcloc}
\end{eqnarray}
where $R_\ast$ represents the extremal point under $c\ll 1$, around which the wavefunction localizes. 
In order to obtain the localized wavefunction around the tip of cone, we require $R_\ast \ll \sqrt{cL^{3/2}-c^2}$, i.e., 
\begin{eqnarray}
1 < \sqrt{cL^{3/2}-c^2} ,
\label{eq:vgen}
\end{eqnarray}
which exhibits a validity condition of the vector mode in the local conifold region. 
As a result, there exist single internal vector zero-mode with $(p,q)=(0,0)$ in the local conifold region only if Eq.~(\ref{eq:vgen}) and $c \ll 1$ are satisfied.

\subsection{Fermions}
\label{subsec:fermions}
In order to describe the Dirac equations, we write down the 
induced D$7$-brane metric $ds^2_{D7}=e_1^2+e_2^2+e_3^2+e_4^2$ 
with respect to the vierbein bases, 
\begin{eqnarray}
e_1&=&\frac{1}{\sqrt{6(C_1^2+C_2^2)}}(C_2d\theta_1 -C_1d\theta_2),
\nonumber\\
e_2&=&\sqrt{\frac{1}{9}+\frac{1}{6(C_1^2+C_2^2)}}(C_1d\theta_1 +C_2d\theta_2),
\nonumber\\
e_3&=&\frac{1}{\sqrt{6(C_1^2+C_2^2)}}(C_2\sin\theta_1 d\phi_1 -C_1\sin\theta_2 d\phi_2),
\nonumber\\
e_4&=&\sqrt{\frac{1}{9}+\frac{1}{6(C_1^2+C_2^2)}}(C_1\sin\theta_1 d\phi_1 +C_2\sin\theta_2 d\phi_2),
\end{eqnarray}
and they are also written in terms of the 
holomorphic coordinates,
\begin{eqnarray}
e_1 &=&-\frac{|z_3||z_4|c}{\sqrt{6\rho}}\left[ \frac{\bar{z}_3dz_3+z_3d\bar{z}_3}{|z_3|^2(|z_3|^2+c^2)}-\frac{\bar{z}_4dz_4+z_4d\bar{z}_4}{|z_4|^2(|z_4|^2+c^2)}\right],
\nonumber\\
e_2 &=&-i\frac{|z_3||z_4|c}{\sqrt{6\rho}}\left[ \frac{\bar{z}_3dz_3-z_3d\bar{z}_3}{|z_3|^2(|z_3|^2+c^2)}-\frac{\bar{z}_4dz_4-z_4d\bar{z}_4}{|z_4|^2(|z_4|^2+c^2)}\right],
\nonumber\\
e_3&=&-\sqrt{\frac{1}{9}+\frac{c^2}{6\rho}}\left[ \frac{\bar{z}_3dz_3+z_3d\bar{z}_3}{|z_3|^2+c^2}-\frac{\bar{z}_4dz_4+z_4d\bar{z}_4}{|z_4|^2+c^2}\right],
\nonumber\\
e_4&=&-i\sqrt{\frac{1}{9}+\frac{c^2}{6\rho}}\left[ \frac{\bar{z}_3dz_3-z_3d\bar{z}_3}{|z_3|^2+c^2}-\frac{\bar{z}_4dz_4-z_4d\bar{z}_4}{|z_4|^2+c^2}\right],
\end{eqnarray}
where $\rho =|z_3|^2+|z_4|^2$. We denote the coefficients of 
these bases, that is, the vierbein $e_{\alpha m}$ itself with the subscripts like $\alpha=1,2,3,4$ and $m=z_3,\bar{z}_3,z_4,\bar{z}_4$. 
In this subsection, the Greek indices represent the local Lorenz frame, while 
the Roman indices label the complex coordinates of the D$7$-brane worldvolume in extra dimensions. 
The vierbeins in the dual basis are defined as 
$\hat{e}_\alpha =e^m_\alpha \partial_m$ and then we 
obtain 
\begin{eqnarray}
\hat{e_1}&=&-\frac{3|z_3||z_4|}{c\sqrt{6\rho}}\left[ \frac{|z_3|^2+c^2}{|z_3|^2}(z_3\partial_{z_3}+\bar{z}_3\partial_{\bar{z}_3})-\frac{|z_4|^2+c^2}{|z_4|^2}(z_4\partial_{z_4}+\bar{z}_4\partial_{\bar{z}_4}) \right],
\nonumber\\ 
\hat{e_2}&=&i\frac{3|z_3||z_4|}{c\sqrt{6\rho}}\left[ \frac{|z_3|^2+c^2}{|z_3|^2}(z_3\partial_{z_3}-\bar{z}_3\partial_{\bar{z}_3})-\frac{|z_4|^2+c^2}{|z_4|^2}(z_4\partial_{z_4}-\bar{z}_4\partial_{\bar{z}_4}) \right],
\nonumber\\ 
\hat{e_3}&=&-\frac{1}{2\rho \sqrt{\frac{1}{9}+\frac{c^2}{6\rho}}}\left[ (|z_3|^2+c^2)(z_3\partial_{z_3}+\bar{z}_3\partial_{\bar{z}_3})+(|z_4|^2+c^2)(z_4\partial_{z_4}+\bar{z}_4\partial_{\bar{z}_4})\right],
\nonumber\\ 
\hat{e_4}&=&i\frac{1}{2\rho \sqrt{\frac{1}{9}+\frac{c^2}{6\rho}}}\left[ (|z_3|^2+c^2)(z_3\partial_{z_3}-\bar{z}_3\partial_{\bar{z}_3})+(|z_4|^2+c^2)(z_4\partial_{z_4}-\bar{z}_4\partial_{\bar{z}_4})\right].
\end{eqnarray}

The zero-mode Dirac equation for the spinor field $\Psi$ on the D$7$-brane includes spin connections when the background geometry has a non-zero curvature,
\begin{eqnarray}
ie_{\nu}^m\tilde{\gamma^\nu}\left( \partial_m +\frac{1}{8}[\tilde{\gamma^\alpha},\tilde{\gamma^\beta}]w_{m\alpha\beta}-iA_m \right) \Psi =0,
\label{eq:Dirac0}
\end{eqnarray}
where $\Psi = ((\psi_1^+) , (\psi_2^-)) \equiv (\Psi_1,\Psi_2,\Psi_3,\Psi_4)$ with the extra-dimensional part of Majorana-Weyl fermions $\psi_1^+$ and $ \psi_2^-$ defined in Eq.~(\ref{eq:chirality}). By imposing the Majorana condition $\lambda^* =B\lambda$ $(B=\Gamma^2 \Gamma^4 \Gamma^7 \Gamma^9)$, $\lambda_3$ ($\lambda_4$) is written in terms of $\lambda_2$ ($\lambda_1$). Thus, we solve the Dirac equation~(\ref{eq:Dirac0}) for $\psi_1^+$ and $\psi_2^-$ and then, $\psi_3^-$ and $\psi_4^+$ are calculated by them.
The explicit form of spin connections are summarized in Appendix~B. 
Since the ten-dimensional gamma matrices are taken as Eq.~(\ref{eq:10Dgamma}), 
the Dirac equation~(\ref{eq:Dirac0}) decomposed on this basis is explicitly rewritten as
\begin{eqnarray}
ie_{\nu}^m\tilde{\gamma^\nu}\left( \partial_m +\frac{1}{8}[\tilde{\gamma^\alpha},\tilde{\gamma^\beta}]w_{m\alpha\beta}-iA_m \right) \Psi=\left(
\begin{array}{cccc}
0 &0&\mathcal{D}_{+11}&\mathcal{D}_{+12}\\
0&0&\mathcal{D}_{+21}&\mathcal{D}_{+22}\\
\mathcal{D}_{-11}&\mathcal{D}_{-12}&0&0\\
\mathcal{D}_{-21}&\mathcal{D}_{-22}&0&0
\end{array}
\right) \left(
\begin{array}{c}
\Psi_1 \\
\Psi_2 \\
\Psi_3 \\
\Psi_4
\end{array}
\right),
\label{eq:Dirac}
\end{eqnarray}
which leads to 
the following simultaneous differential equations,
\begin{eqnarray}
\mathcal{D}_{+11}\Psi_3 +\mathcal{D}_{+12}\Psi_4 &=&0,
\nonumber\\
\mathcal{D}_{+21}\Psi_3 +\mathcal{D}_{+22}\Psi_4 &=&0,
\nonumber\\
\mathcal{D}_{-11}\Psi_1 +\mathcal{D}_{-12}\Psi_2 &=&0,
\nonumber\\
\mathcal{D}_{-21}\Psi_1 +\mathcal{D}_{-22}\Psi_2 &=&0.
\label{eq:diraccom}
\end{eqnarray}
In the following, we concentrate on the upper right elements 
of the Dirac operator acting on $\Psi_3$ and $\Psi_4$ in Eq.~(\ref{eq:Dirac}), 
those are explicitly written as 
\begin{eqnarray}
\mathcal{D}_{+11}&=&-{\cal A} \left[ \partial_{z_3} +\frac{M_1}{z_3}\ln(|z_3|^2+c^2)\right] +{\cal B}\left[  \partial_{z_4} +\frac{M_2}{z_4}\ln(|z_4|^2+c^2) \right] -{\cal E}+M',
\nonumber\\
\mathcal{D}_{+12}&=&-i{\cal C} \left[ \partial_{\bar{z}_3} -\frac{M_1}{\bar{z}_3}\ln(|z_3|^2+c^2)\right] -i{\cal D}\left[  \partial_{\bar{z}_4} -\frac{M_2}{\bar{z}_4}\ln(|z_4|^2+c^2) \right] +i{\cal F},
\nonumber\\
\mathcal{D}_{+21}&=&-i\bar{{\cal C}} \left[ \partial_{z_3} +\frac{M_1}{z_3}\ln(|z_3|^2+c^2)\right] -i\bar{{\cal D}}\left[  \partial_{z_4} +\frac{M_2}{z_4}\ln(|z_4|^2+c^2) \right] +i{\cal F},
\nonumber\\
\mathcal{D}_{+22}&=&-\bar{{\cal A}} \left[ \partial_{\bar{z}_3} -\frac{M_1}{\bar{z}_3}\ln(|z_3|^2+c^2)\right] +\bar{{\cal B}}\left[  \partial_{\bar{z}_4} -\frac{M_2}{\bar{z}_4}\ln(|z_4|^2+c^2) \right] -{\cal E}-M',
\end{eqnarray}
where
\begin{align}
{\cal A}&=\frac{6|z_3||z_4|(|z_3|^2+c^2)}{c\sqrt{6\rho}\bar{z}_3},
\qquad
{\cal B}=\frac{6|z_3||z_4|(|z_4|^2+c^2)}{c\sqrt{6\rho}\bar{z}_4},
\nonumber\\
{\cal C}&=\frac{|z_3|^2(|z_3|^2+c^2)}{\rho \sqrt{\frac{1}{9}+\frac{c^2}{6\rho}}z_3},
\qquad
{\cal D}=\frac{|z_4|^2(|z_4|^2+c^2)}{\rho \sqrt{\frac{1}{9}+\frac{c^2}{6\rho}}z_4},
\nonumber\\
{\cal E}&=\frac{3\Bigl( |z_3|^2|z_4|^2(|z_3|^2-|z_4|^2)+c^2(|z_4|^4-|z_3|^4) \Bigl) }{2\rho c\sqrt{6\rho} |z_3||z_4|}+\frac{|z_3||z_4|c\,\Bigl( |z_4|^2-|z_3|^2 \Bigl) }{4\rho^2 \sqrt{6\rho}\left( \frac{1}{9}+\frac{c^2}{6\rho} \right)},
\nonumber\\
{\cal F}&=\frac{ 2|z_3|^4+10|z_3|^2|z_4|^2+2|z_4|^4 -3\rho c^2 }{12\rho^2  \sqrt{\frac{1}{9}+\frac{c^2}{6\rho}}} -\frac{c^2\,\Bigl( \rho^2 +6|z_3|^2|z_4|^2+3\rho c^2\Bigl) }{72\rho^3 \left( \frac{1}{9}+\frac{c^2}{6\rho} \right)^{3/2}},
\nonumber\\
M'&=\frac{3 c^2}{|z_3||z_4|c\sqrt{6\rho}} \left( M_1 |z_4|^2 -M_2 |z_3|^2 \right).
\end{align}

By comparing Eq.~(\ref{eq:diraccom}) with those led to Eq.~(\ref{eq:bosonwave}) in the bosonic case, 
the following forms of the fermionic wavefunctions $\Psi_3$ and $\Psi_4$ are expected:
\begin{align}
\Psi_3 =F(z_3,z_4,\bar{z}_3,\bar{z}_4)&e^{M_1\left( -\ln c^2 \ln z_3 +{\rm Li}_2(-\frac{|z_3|^2}{c^2})\right)}e^{M_2\left( -\ln c^2 \ln z_4 +{\rm Li}_2(-\frac{|z_4|^2}{c^2})\right)}\nonumber\\ 
&\times \left( \frac{z_3}{|z_3|^2+c^2}\right)^{\frac{M_1}{2}}\left( \frac{z_4}{|z_4|^2+c^2}\right)^{\frac{M_2}{2}},
\nonumber\\
\Psi_4 =K(z_3,z_4,\bar{z}_3,\bar{z}_4)&e^{-M_1\left( -\ln c^2 \ln \bar{z}_3 +{\rm Li}_2(-\frac{|z_3|^2}{c^2})\right)}e^{-M_2\left( -\ln c^2 \ln \bar{z}_4 +{\rm Li}_2(-\frac{|z_4|^2}{c^2})\right)} \nonumber\\
&\times \left( \frac{\bar{z}_3}{|z_3|^2+c^2}\right)^{-\frac{M_1}{2}}\left( \frac{\bar{z}_4}{|z_4|^2+c^2}\right)^{-\frac{M_2}{2}},
\label{eq:psi341}
\end{align}
where $F(z_3,z_4,\bar{z}_3,\bar{z}_4)$ and $K(z_3,z_4,\bar{z}_3,\bar{z}_4)$ are functions whose holomorphic and anti-holomorphic part are fixed respectively as follows. By substituting Eq.~(\ref{eq:psi341}) into 
Eq.~(\ref{eq:diraccom}), we obtain
\begin{eqnarray}
\left[ {\cal A}\frac{\partial_{z_3} F}{F}-{\cal B}\frac{\partial_{z_4} F}{F}+{\cal E} \right]\Psi_3 +i\left[ {\cal C}\frac{\partial_{\bar{z}_3}K}{K}+{\cal D}\frac{\partial_{\bar{z}_4}K}{K}-{\cal F} \right] \Psi_4 &=&0,
\nonumber\\
i\left[ \bar{{\cal C}}\frac{\partial_{z_3}F}{F}+\bar{{\cal D}}\frac{\partial_{z_4}F}{F}-{\cal F} \right] \Psi_3 +\left[ \bar{{\cal A}}\frac{\partial_{\partial{z_3}} K}{K}-\bar{{\cal B}}\frac{\partial_{\bar{z}_4} K}{K}+{\cal E} \right]\Psi_4 &=&0.
\label{eq:dirac342}
\end{eqnarray}
Then we consider such a case that all the coefficients of $\Psi_3$ and $\Psi_4$ vanish independently to each other, that leads to
\begin{eqnarray}
\partial_{z_3}F &=&\left[ \frac{2}{3(|z_3|^2+c^2)}-\frac{1}{4|z_3|^2}-\frac{1}{2(2\rho +3c^2)} \right] \bar{z}_3F,
\nonumber\\
\partial_{z_4}F &=&\left[ \frac{2}{3(|z_4|^2+c^2)}-\frac{1}{4|z_4|^2}-\frac{1}{2(2\rho +3c^2)} \right] \bar{z}_4F,
\nonumber\\
\partial_{\bar{z}_3}K &=&\left[ \frac{2}{3(|z_3|^2+c^2)}-\frac{1}{4|z_3|^2}-\frac{1}{2(2\rho +3c^2)} \right]z_3 K,
\nonumber\\
\partial_{\bar{z}_4}K &=&\left[ \frac{2}{3(|z_4|^2+c^2)}-\frac{1}{4|z_4|^2}-\frac{1}{2(2\rho +3c^2)} \right]z_4 K.
\label{eq:diracdecom}
\end{eqnarray}
The functions $F$ and $K$ satisfying Eq.~(\ref{eq:diracdecom}) can be described as 
\begin{eqnarray}
F(z_3,z_4,\bar{z}_3,\bar{z}_4)&=& G(\bar{z}_3,\bar{z}_4)(|z_3|^2+c^2)^{\frac{2}{3}}(|z_4|^2+c^2)^{\frac{2}{3}}(2\rho +3c^2)^{-\frac{1}{4}}z_3^{-\frac{1}{4}}z_4^{-\frac{1}{4}},
\nonumber\\
K(z_3,z_4,\bar{z}_3,\bar{z}_4)&=&H(z_3,z_4)(|z_3|^2+c^2)^{\frac{2}{3}}(|z_4|^2+c^2)^{\frac{2}{3}}(2\rho +3c^2)^{-\frac{1}{4}}\bar{z}_3^{-\frac{1}{4}}\bar{z}_4^{-\frac{1}{4}},
\end{eqnarray}
where $G(\bar{z}_3,\bar{z}_4)$ and $H(z_3,z_4)$ are anti-holomorphic and holomorphic functions, constrained by the normalization conditions of fermions. 
In the same way as the bosonic one, we impose their form as $G(\bar{z}_3,\bar{z}_4)=\bar{z}_3^a \bar{z}_4^b$ and $ H(z_3,z_4)=z_3^az_4^b$. 

Finally, the explicit forms of the fermionic wavefunctions are 
\begin{align}
\Psi_3 &=G(\bar{z}_3,\bar{z}_4)e^{M_1\left( -\ln c^2 \ln z_3 +{\rm Li}_2(-\frac{|z_3|^2}{c^2})\right)}e^{M_2\left( -\ln c^2 \ln z_4 +{\rm Li}_2(-\frac{|z_4|^2}{c^2})\right)}\left( \frac{z_3}{|z_3|^2+c^2}\right)^{\frac{M_1}{2}}\left( \frac{z_4}{|z_4|^2+c^2}\right)^{\frac{M_2}{2}}\nonumber\\
&\times (|z_3|^2+c^2)^{\frac{2}{3}}(|z_4|^2+c^2)^{\frac{2}{3}}(2\rho +3c^2)^{-\frac{1}{4}}z_3^{-\frac{1}{4}}z_4^{-\frac{1}{4}},
\nonumber\\
\Psi_4 &=H(z_3,z_4)e^{-M_1\left( -\ln c^2 \ln \bar{z}_3 +{\rm Li}_2(-\frac{|z_3|^2}{c^2})\right)}e^{-M_2\left( -\ln c^2 \ln \bar{z}_4 +{\rm Li}_2(-\frac{|z_4|^2}{c^2})\right)}\left( \frac{\bar{z}_3}{|z_3|^2+c^2}\right)^{-\frac{M_1}{2}}\left( \frac{\bar{z}_4}{|z_4|^2+c^2}\right)^{-\frac{M_2}{2}}\nonumber\\
&\times (|z_3|^2+c^2)^{\frac{2}{3}}(|z_4|^2+c^2)^{\frac{2}{3}}(2\rho +3c^2)^{-\frac{1}{4}}\bar{z}_3^{-\frac{1}{4}}\bar{z}_4^{-\frac{1}{4}},
\label{eq:fermionwave}
\end{align}
whereas $\Psi_1$ and $\Psi_2$ have no solution satisfying 
Eq.~(\ref{eq:dirac342}). 

When we impose the periodic boundary condition $\Psi_{3,4} (\theta_i, \phi_i +2\pi)=\Psi_{3,4} (\theta_i, \phi_i)$, wavefunctions are proportional to the integer power of $z_i$ and/or $\bar{z}_i$. 
Therefore, the arbitrary functions in Eq.~(\ref{eq:fermionwave}) are constrained as
\begin{eqnarray}
G(\bar{z}_3,\bar{z}_4)=\sum_{a, b \in {\mathbb Z}} A_{a,b}^G \bar{z}_3^{a+M_1(-\ln c^2 +1/2)-1/4} \bar{z}_4^{b+M_2(-\ln c^2 +1/2)-1/4},\nonumber\\
H(z_3,z_4)=\sum_{a', b' \in {\mathbb Z}} A_{a',b'}^H z_3^{a'-M_1(-\ln c^2 +1/2)-1/4} z_4^{b'-M_2(-\ln c^2 +1/2)-1/4},
\label{eq:Fourierfermion}
\end{eqnarray}
where the integers $a$, $a'$, $b$, $b'$ and real constants $A_{a,b}^{G,H}$ are determined by the normalization condition in the same way as the bosonic wavefunction. 
By substituting Eq.~(\ref{eq:Fourierfermion}) to Eq.~(\ref{eq:fermionwave}), each mode of the zero-mode fermionic wavefunctions is rewritten to be
\begin{eqnarray}
\Psi_{3,ab}=A_{a,b}^G e^{M_1 {\rm Li}_2 (-\frac{|z_3|^2}{c^2})}e^{M_2 {\rm Li}_2 (-\frac{|z_4|^2}{c^2})}(|z_3|^2+c^2)^{-\frac{M_1}{2}+\frac{2}{3}}(|z_4|^2+c^2)^{-\frac{M_2}{2}+\frac{2}{3}}(2\rho +3c^2)^{-\frac{1}{4}}\nonumber\\
\times |z_3|^{2M_1(-\ln c^2 +1/2)-\frac{1}{2}}|z_4|^{2M_2(-\ln c^2 +1/2)-\frac{1}{2}}\bar{z}_3^a \bar{z}_4^b,\nonumber\\
\Psi_{4,a'b'}=A_{a',b'}^H e^{-M_1 {\rm Li}_2 (-\frac{|z_3|^2}{c^2})}e^{-M_2 {\rm Li}_2 (-\frac{|z_4|^2}{c^2})}(|z_3|^2+c^2)^{\frac{M_1}{2}+\frac{2}{3}}(|z_4|^2+c^2)^{\frac{M_2}{2}+\frac{2}{3}}(2\rho +3c^2)^{-\frac{1}{4}}\nonumber\\
\times |z_3|^{-2M_1(-\ln c^2 +1/2)-\frac{1}{2}}|z_4|^{-2M_2(-\ln c^2 +1/2)-\frac{1}{2}}z_3^{a'} z_4^{b'},
\label{eq:fw}
\end{eqnarray}
Since the different modes of wavefunctions $\Psi_{3,ab}$ ($\Psi_{4,a'b'}$) are orthogonal to each other, i.e., 
$\int dz_3^2dz_4^2 \sqrt{g} \Psi_{3,a_1b_1}^\dagger \Psi_{3,a_2b_2}=0$ for $a_1\neq a_2$ or $b_1\neq b_2$, (similarly for $\Psi_{4,a'b'}$) due to the periodicity, they are independent solutions for the Dirac equation~(\ref{eq:Dirac}) labelled by the integers $a$, $a'$, $b$ and $b'$. 
We will see below that the normalization and validity conditions of wavefunctions in the local conifold region restrict the allowed integers for $a$, $a'$, $b$ and $b'$ as in the bosonic case, then the number of possible combinations of them corresponds to the degeneracy of zero-modes, which is identified as the number of fermionic generations.
We estimate 
the normalization conditions in three cases, the vanishing fluxes $M_1=M_2=0$, the supersymmetric fluxes $M_1=-M_2 \equiv M \ne 0$ and the non-supersymmetric fluxes $M_1=M_2 \equiv M \ne 0$ in the following.

\medskip

\noindent
$\bullet \quad M_1=-M_2 \equiv M \ne 0$

\medskip

The fermionic wavefunctions with the supersymmetric fluxes are non-normalizable as it is in the bosonic case, 
which can be understood as follows. 
The normalization condition of $\Psi_{3,ab}$ in this case is given by 
\begin{eqnarray}
\int d^2 z_3 d^2 z_4 \sqrt{g}\Psi_{3,a_ib_i}^\dagger \Psi_{3,a_jb_j} = \delta_{a_ia_j}\delta_{b_ib_j},
\end{eqnarray}
which is explicitly calculated as
\begin{eqnarray}
1=
\frac{16\pi^2 c^2}{27}(A_{ab}^G)^2\int dr_3 dr_4
e^{2M{\rm Li}_2(-\frac{r_3^2}{c^2})}e^{-2M{\rm Li}_2(-\frac{r_4^2}{c^2})}
 \frac{(2r_3^2+2r_4^2+3c^2)^{1/2}}{ (r_3^2+c^2)^{\frac{2}{3}+M} (r_4^2+c^2)^{\frac{2}{3}-M}}\nonumber\\
\times  r_3^{2a+4M(-\ln c^2 +1/2)} r_4^{2b-4M(-\ln c^2 +1/2)}.
\label{eq:MferN}
\end{eqnarray}
From the asymptotic expansion of the dilogarithm function in the limit $r \gg 1$, 
\begin{eqnarray}
{\rm Li}_2\left(-\frac{r^2}{c^2}\right)\simeq -\frac{\pi^2}{6}-\frac{1}{2}\left(\ln \left(\frac{r^2}{c^2}\right)\right)^2 +{\cal O}\left(\frac{c^2}{r^2}\right),
\end{eqnarray}
the factor $e^{-2M{\rm Li}_2(-r_4^2/c^2)}$ ($e^{2M{\rm Li}_2(-r_3^2/c^2)}$) in the integrand of Eq~.(\ref{eq:MferN}) diverges in the limit $r_4\gg c$ ($r_3\gg c$) if the flux $M$ is chosen as positive (negative) value. 
Therefore we find that the fermionic wavefunction $\Psi_{3,ab}$ with the supersymmetric fluxes is also non-normalizable.
%although it is normalizable for the finite ranges 
%$z_3,z_4\in [0,z_{max}]$. 
A similar analysis for $\Psi_{4,a'b'}$ leads to the same results as $\Psi_{3,ab}$. 

\medskip

\noindent
$\bullet \quad M_1=M_2 \equiv M \ne 0$

\medskip

The non-supersymmetric choice of fluxes~(\ref{eq:FI}) gives rise to flip $M_2$ in the supersymmetric one to $-M_2$. 
It causes the fermionic wavefunction $\Psi_{3,ab}$ ($\Psi_{4,a'b'}$) to be a normalizable one when the flux~$M$ is positive (negative) due to the nature of dilogarithm function.
Here, we assume that the flux induced FI-term is canceled by some VEVs of charged scalar fields as mentioned before. 

Also in the analysis of fermionic wavefunctions, it has to be taken into account the validity of them in local conifold region inside the boundary of near horizon limit discussed in Sec.~\ref{subsec:bosons}. 
Without loss of generality, we focus on $\Psi_{3,ab}$ with the non-supersymmetric fluxes, which is asymptotically given in the limit $R=r_3=r_4\gg c$ as 
\begin{eqnarray}
\hat{\Psi}_{3,ab}\equiv 4\pi R(\sqrt{g})^{1/2}|\Psi_{3}|&\propto& R^{a+b-4M\ln c^2-\frac{1}{3}}e^{-M(\ln (-R^2/c^2))^2 } \nonumber\\
&=& R^{a+b-2M\ln (c^2 R^2)-\frac{1}{3}}c^{2M\ln (R^2/c^2)},
\end{eqnarray}
by employing Eq.~(\ref{eq:dialog}) and then the extremal condition of $\hat{\Psi}_{3,ab}$ is achieved by the following relation,
\begin{eqnarray}
a+b-\frac{1}{3}-8M\ln R_{\ast}=0,
\label{eq:ferloc}
\end{eqnarray}
where $R_\ast$ is the extremal point. 
In order to obtain the localized wavefunction around the tip of cone, we require $R_\ast \ll \sqrt{cL^{3/2}-c^2}$, that is, 
\begin{eqnarray}
a+b < 8M\ln \sqrt{cL^{3/2}-c^2} +\frac{1}{3},
\label{eq:genfer}
\end{eqnarray}
which represents a validity condition of the fermionic wavefunction in the local conifold region. 
In the same way as bosonic case, the lower bounds of $a$ and $b$ are determined as
\begin{eqnarray}
a>-2M(-\ln c^2 +\frac{1}{2}),\quad b>-2M(-\ln c^2 +\frac{1}{2}). 
\label{eq:genfer2}
\end{eqnarray}

\begin{figure}
\centering \leavevmode
\includegraphics[width=90mm]{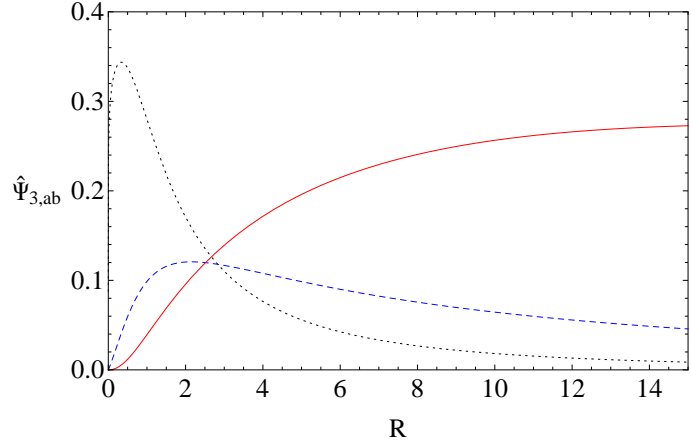}
\caption{The fermionic wavefunction $\hat{\Psi}_{3,ab}$ with nonzero 
FI term and $N_1=N_2=1$ on the $R=r_3=r_4$ hypersurface. 
The black dotted ($a=b=0$) and the blue dashed ($a=0,b=1$) curves are localized at the point below the horizon scale $R\simeq 7$ in contrast to the red solid curve ($a=1,b=1$). The curve parameterized in $a=1,b=0$ is the same as the blue dashed curve ($a=0,b=1$).}
\label{fig:bs_FI2}
\end{figure}

With the parameters given by 
\begin{eqnarray}
c=1,\,\,\, N=1,\,\,\, L\simeq 13.5,\,\,\, M_s=1.08\times 10^{17}\,{\rm GeV},
\label{eq:para3}
\end{eqnarray}
the fermionic wavefunction $\hat{\Psi}_{3,ab}$ on the $R=r_3=r_4$ hypersurface are drawn in Fig.~\ref{fig:bs_FI2} by computing normalization factors in a numerical way. 
In Fig.~\ref{fig:bs_FI2}, the wavefunction drawn by the red-solid curve is the one out of control, because it localizes outside the boundary of near horizon limit, $R_\ast>\sqrt{cL^{3/2}-c^2}\simeq 7$. 
Such an observation is consistent with the above argument leading to Eq.~(\ref{eq:ferloc}). 
In Fig.~\ref{fig:bs_FI2}, we find three independent zero-mode solutions in this case, those correspond to $(a,b)=(0,0),(1,0)$ and $(0,1)$. 
A similar analysis for the other wavefunction $\Psi_{4,a'b'}$ but with the negative flux $M<0$ can be performed, and the result is the same as $\Psi_{3,ab}$. 

\newpage
\noindent
$\bullet \quad M_1=M_2=0$

\medskip

In the case of vanishing fluxes, the normalization condition of $\Psi_{3,ab}$ is expressed as 
\begin{align}
1&=
\frac{16\pi^2 c^2}{27}(A_{a,b}^G)^2\int dr_3 dr_4 (r_3^2+c^2)^{-\frac{2}{3}}(r_4^2+c^2)^{-\frac{2}{3}}r_3^{2a}r_4^{2b}(2r_3^2+2r_4^2+3c^2)^{\frac{1}{2}}
\nonumber\\
&=\frac{2\sqrt{2}\pi^2 c^{\frac{2}{3}}}{27} (A_{a,b}^G)^2
\int dr_3 \frac{r_3^{2a} }{(r_3^2+c^2)^{\frac{2}{3}}} \biggl[ 
 -\frac{4c^{2+2b}{\rm csc}(b\pi) \Gamma (-b-\frac{1}{3} )}{\Gamma (\frac{2}{3}) \Gamma (-b)}
 {}_2F_1 \left( -\frac{1}{2},-b-\frac{1}{3},-b;\frac{r_3^2}{c^2} +\frac{3}{2} \right) \nonumber\\
&\hspace{4cm}+ 8^{-b}\frac{(2r_3^2+3c^2)^{1+b} \Gamma (-b) \Gamma (2b+1)}{\Gamma (b+2)} 
{}_2F_1 \left( \frac{2}{3},b+\frac{1}{2},b+2;\frac{r_3^2}{c^2} +\frac{3}{2} \right)
\biggl],
\end{align}
where the integral over the infinite region as in the bosonic case is convergent only if $-\frac{1}{2}<a,b<-\frac{1}{3}$ are satisfied. 
As a result, these range do not include integers $a$ and $b$, and there are no normalizable modes in this case.
A similar analysis for $\Psi_{4,a'b'}$ leads to the same result as $\Psi_{3,ab}$. 

As we discussed in the case of non-supersymmetric flux, 
taking the validity condition~(\ref{eq:genfer}) and the convergence condition of the normalization factors~(\ref{eq:genfer2}) into account, there are no zero-mode solutions with integers $a$ and $b$. 
However, by introducing the effect of twisting, the situation changes. 
Actually, by shifting the number of fluxes, at least one fermionic zero-mode is allowed by conditions~(\ref{eq:genfer}) and~(\ref{eq:genfer2}). In particular, there appears a zero-mode solution with $(a,b)=(0,0)$ for the minimal shift $N\rightarrow N+1/2$ where the half-integer twist is demanded by double-valuedness of the spinor.
\vspace{5pt}

\section{Phenomenological aspects}
\label{sec:AdS5}
In this section, we show some phenomenological aspects of localized zero-modes, derived in the previous section, in the low-energy effective theories with five and four spacetime dimensions.

Before entering upon a discussion of the detail, let us comment on the correspondence between obtained zero-modes in the previous section and supersymmery.
In the local conifold region, $4$D ${\cal N}=1$ supermultiplets are categorized into single vector multiplet ${\bf V}=\{ A_\mu, \eta_2\}$ and triple chiral multiplets ${\bf \Phi}_i =\{ A_i, \eta_i\}$ ($i=1,3,4$) where $\eta_2$ ($\eta_i$) comes from a component of 10D Majorana-Weyl fermion $\lambda_2$ ($\lambda_i$). Here, $\lambda_{1,4}$ sector corresponds to the fermionic partners of internal vector modes and $\lambda_{2,3}$ sector is identified as the partner of transverse scalars and gauge bosons of 4D SYM, respectively.
From the result of Sec.~\ref{sec:3} with the ansatz~(\ref{eq:para1}) and~(\ref{eq:para2}) without a flux, there are three generations for the transverse scalar, single Minkowski vectors and no solutions for the internal vectors. In order to achieve supersymmetric spectrum, four fermionic zero-modes have to be generated, for example, three fermions $\Psi_3$ with the twist $N\rightarrow N+1$ and a single fermion $\Psi_4$ with the twist $N\rightarrow N-1/2$. As mensioned before, we do not identify the detail of twisting, because it could be further modified when the conifold is embedded into a global CY manifold. 
In any case, the twist can cause a mixture of bases for the Majorana-Weyl fermions $\Psi_i$ in 4D ${\cal N}=1$ supermultiplets. In such a case, supersymmetric models which contain the fermionic partners of internal vector modes would be also possible, where the Yukawa interaction terms are yielded in the form of $\lambda_1 \lambda_1 \phi$ for the transverse modes $\phi$ or $\lambda_1 \lambda_2 A_m$ for the internal vector modes $A_m$. For more details, see Ref.~\cite{Conlon:2008qi} and references therein. In Sec.~\ref{subsec:Yukawa}, we will show the particular example of the triple overlap integrals which would appear in such Yukawa interaction terms.
\subsection{Wavefucntions on $AdS_5$}
First, we discuss the properties of matter wavefunction from the viewpoint of five-dimensional (5D) effective theory. 
In the near-horizon limit, the effective $5$D background metric is extracted as 
\begin{align}
ds_5^2&=\frac{r^2}{L^2} d^2x_{1,3} + \frac{L^2}{r^2} dr^2,
\end{align}
which is also rewritten in terms of  $r=Le^{-y/L}$,
\begin{align}
ds_5^2= e^{-2y/L}d^2x_{1,3} +dy^2,
\label{eq:5Dmetric}
\end{align}
where $L$ corresponds to the inverse of $AdS_5$ curvature. 
As pointed out in Refs.~\cite{Brummer:2005sh}, the our set-up is similar to those of Randall-Sundrum like model~\cite{Randall:1999ee} 
where the IR brane is located at the tip of conifold, while the Planck brane is included in 
the remaining Calabi-Yau manifold. 
In our setup, the locations of the IR brane or the Planck brane for $AdS_5$ coordinate are taken as $y_H$ or $y_{\rm pl}$ respectively, 
and correspondingly, $r_{\rm min}$ and $r_{\rm max}$ can be rewritten by $y_{\rm max}= -L\ln \left( c^{2/3}/L\right)$ and $y_{\rm min}= -2L\ln L$.
Note that the matter fields localize towards the $y_{\max}$ rather than the tip of conifold. 
On the other hand, the Planck brane can be located at any points including $y_{\rm pl}<y_{\min}$ under the assumption that 
the structure of $AdS_5$ space also holds out of the near horizon region.

As shown in Secs.~\ref{subsec:bosons} and \ref{subsec:fermions}, the obtained wavefunctions of bosons and fermions 
are normalizable with non-supersymmetric fluxes. 
Figs.~\ref{fig:bs_FI} and \ref{fig:bs_FI2} show that these wavefunctions on the hypersurface $|z_3|=|z_4|$ localize towards the tip of conifold. 
In such a case, they are also rewritten in terms of the coordinate $y$, 
\begin{align}
\Phi_{13,pq}&\propto e^{2M{\rm Li}_2\left( 1-\frac{L^{3/2}}{c} \exp\left( -\frac{3y}{2L}\right)\right) +\frac{3y}{2L}M}\left( L^{\frac{3}{2}}e^{-\frac{3y}{2L}}-c \right)^{2M\left( -\ln c^2 +\frac{1}{2}\right)+\frac{1}{2}(p+q)},
\nonumber\\
\Phi_{24,pq}&\propto e^{-2M{\rm Li}_2\left( 1-\frac{L^{3/2}}{c} \exp\left( -\frac{3y}{2L}\right)\right) -\frac{3y}{2L}M}\left( L^{\frac{3}{2}}e^{-\frac{3y}{2L}}-c \right)^{-2M\left( -\ln c^2 +\frac{1}{2}\right)+\frac{1}{2}(p+q)},
\label{eq:bosony}
\end{align}
for bosons and 
\begin{align}
\Psi_{3,ab}&\propto e^{2M{\rm Li}_2\left( 1-\frac{L^{3/2}}{c} \exp\left( -\frac{3y}{2L}\right)\right) +\frac{3y}{2L}M-\frac{2y}{L}}\left( 2Le^{-\frac{3y}{2L}}+1\right)^{-\frac{1}{4}} \left( L^{\frac{3}{2}}e^{-\frac{3y}{2L}}-c \right)^{2M\left( -\ln c^2 +\frac{1}{2}\right)+\frac{1}{2}(a+b-1)},
\nonumber\\
\Psi_{4,ab}&\propto e^{-2M{\rm Li}_2\left( 1-\frac{L^{3/2}}{c} \exp\left( -\frac{3y}{2L}\right)\right) -\frac{3y}{2L}M-\frac{2y}{L}}\left( 2Le^{-\frac{3y}{2L}}+1\right)^{-\frac{1}{4}} \left( L^{\frac{3}{2}}e^{-\frac{3y}{2L}}-c \right)^{-2M\left( -\ln c^2 +\frac{1}{2}\right)+\frac{1}{2}(a+b-1)},
\label{eq:fermiony}
\end{align}
for fermions. Eqs.~(\ref{eq:bosony}) and (\ref{eq:fermiony}) show that the wavefunctions of bosons and fermions have the 
exponential form determined by the $U(1)$ flux. 
It is known that exponential profile of wavefunctions for graviton~\cite{Randall:1999ee} and matter~\cite{Chang:1999nh} zero-modes can yield a weak-Planck hierarchy and Yukawa hierarchies of quarks and leptons, respectively, in the framework of $AdS_5$ supergravity~\cite{Altendorfer:2000rr}.
In our case, the exponential forms of bosons and fermions are originated from the 
$U(1)$ magnetic flux which suggests us to obtain some hierarchical structures of overlap integrals 
between boson and fermions as shown in the next subsection. 

Especially, the wavefunction of matter fields in the direction of $AdS_5$ 
are approximately estimated as 
\begin{equation}
\Phi_{13,pq} \propto 
\left \{
\begin{array}{l}
e^{y\frac{3M}{L}\left( \ln (L^{\frac{3}{2}}c)-\frac{1}{4M}(p+q)\right)-y^2\frac{9M}{4L^2}}\hspace{0.6cm} \left( y\ll -L \right),
\nonumber\\
e^{y\frac{9M}{2L}+(M\ln y)(-2\ln c^2 +1+\frac{1}{2M}(p+q))}\hspace{0.2cm} \left( y\simeq y_{\rm max} \right),
\end{array}
\right.
\end{equation}
\begin{equation}
\Phi_{24,pq} \propto 
\left \{
\begin{array}{l}
e^{-y\frac{3M}{L}\left( \ln (L^{\frac{3}{2}}c)+\frac{1}{4M}(p+q)\right)+y^2\frac{9M}{4L^2}}\hspace{0.6cm} \left( y\ll -L \right),
\\
e^{-y\frac{9M}{2L}-(M\ln y)(-2\ln c^2 +1-\frac{1}{2M}(p+q))}\hspace{0.2cm}\left( y\simeq y_{\rm max} \right),
\end{array}
\right.
\label{eq:Bbulk}
\end{equation}
for bosons and 
\begin{equation}
\Psi_{3,ab} \propto 
\left \{
\begin{array}{l}
e^{y\frac{3M}{L}\left( \ln (L^{\frac{3}{2}}c)-\frac{1}{4M}(a+b+\frac{7}{6})\right)-y^2\frac{9M}{4L^2}}\hspace{1.2cm} \left( y\ll -L \right),
\nonumber\\
e^{y\left( \frac{9M}{2L} -\frac{2}{L}\right)+(M\ln y)(-2\ln c^2 +1+\frac{1}{2M}(a+b-1))}\left( y\simeq y_{\rm max} \right),
\end{array}
\right.
\end{equation}
\begin{equation}
\Psi_{4,ab} \propto 
\left \{
\begin{array}{l}
e^{-y\frac{3M}{L}\left( \ln (L^{\frac{3}{2}}c)+\frac{1}{4M}(a+b+\frac{7}{6})\right)+y^2\frac{9M}{4L^2}}\hspace{1.2cm} \left( y\ll -L \right),
\\
e^{-y\left( \frac{9M}{2L} +\frac{2}{L}\right)-(M\ln y)(-2\ln c^2 +1-\frac{1}{2M}(a+b-1))}\left( y\simeq y_{\rm max} \right),
\end{array}
\right.
\label{eq:Fbulk}
\end{equation}
for fermions. 
Although, in the effective $AdS_5$ supergravity~\cite{Altendorfer:2000rr} the bulk masses of the matter fields are proportional to their graviphoton charges which are free parameters, in our setup, we identify the origin of such parameters as the generation numbers associated with the $U(1)$ flux as can be seen in 
Eqs.~(\ref{eq:Bbulk}) and~(\ref{eq:Fbulk}). 

Next let us consider an example which would solve the gauge hierarchy problem where the matter fields, especially, the Higgs fields localize around $y=y_H$ due to a suitable choice of 
magnetic fluxes, 
while the graviton localize towards the CY manifold, characterized by $y=y_{\rm pl}$. 
As shown in Ref.~\cite{Randall:1999ee}, 
when the VEVs of $5$D original and $4$D effective Higgs fields are denoted by 
$v_0$ and $v$, respectively, they are related as 
\begin{eqnarray}
v=e^{-\frac{y_H-y_{\rm pl}}{L}}v_0.
\end{eqnarray}
Although, in the case with $y_{\rm pl}=y_{\rm min}$ and $y_H=y_{\rm max}$, the Higgs VEVs between $y_H$ and $y_{\rm pl}$ are not so 
suppressed such as $v\simeq (c^{2/3}v_0)/L^3$, the hierarchy between electroweak (EW) and Planck scale can be explained if $y_{\rm pl}$ takes more smaller values under the assumption that the effective $AdS_5$ description is valid even outside the near horizon limit.\footnote{In other words, we implicitly consider such a global CY space that the assumption here is valid.}
It is remarkable that these features are determined by the localization profile of matter fields around the tip of conifold.

\subsection{Overlap integrals}
\label{subsec:Yukawa}
%\begin{figure}[t,b]
%\centering \leavevmode
%\includegraphics[width=90mm]{bosonnosusy.eps}
%\caption{The wavefunction of transverse scalar mode with nonzero 
%flux $M=1$ and $c=1$.}
%\label{fig:bosonnosusy}
%\end{figure}
In Sec.~\ref{subsec:bosons} and \ref{subsec:fermions}, we have studied the matter wavefunction and its properties. 
Here we consider the overlap integrals among bifundamental fields starting from the $8$D $U(N)$ SYM theory by introducing the following magnetic fluxes, 
\begin{eqnarray}
F_{{\frak a}{\frak b}}=\left(
\begin{array}{ccc}
M_{1(N_1)}\1_{N_1}&&\\
&M_{1(N_2)}\1_{N_2}&\\
&&\ddots
\end{array}
\right)
Q_1 \Omega_{11} +\left(
\begin{array}{ccc}
M_{2(N_1)}\1_{N_1}&&\\
&M_{2(N_2)}\1_{N_2}&\\
&&\ddots
\end{array}
\right)
Q_2 \Omega_{22},
\label{eq:U(N)fluxes}
\end{eqnarray}
where $\1_{N_i}$ is an $N_i \times N_i$ unit matrix. 
These magnetic fluxes break $U(N)$ gauge group as $U(N) \to U(N_1)\times U(N_2)\times \cdots U(N_n)$ with $ \sum_{a=1}^n N_a =N$. 
As mentioned in Sec.~\ref{subsec:red}, each bifundamental field $({\bf N}_s,\bar{\bf N}_t), ((\bar{\bf N}_s, {\bf N}_t)) $ for $s,t=1,\ldots, n, (s\neq t)$ in the 
off-diagonal components of $U(N)$ feels $M_{1(N_s)}-M_{1(N_t)}$ and $M_{2(N_s)}-M_{2(N_t)}$ ( $M_{1(N_t)}-M_{1(N_s)}$ and 
$M_{2(N_t)}-M_{2(N_s)}$) units of fluxes, because the covariant derivative for these fields includes a commutation relation as shown in Eq.~(\ref{eq:covariantD}).

In the following, we discuss about overlap integrals among a single boson and two fermion wavefunctions, those might be involved in the Yukawa interaction in the $4$D effective theory or some operators beyond the leading SYM approximation.

The overlap integrals of extra-dimensional wavefunctions are given by 
\begin{eqnarray}
Y_{(a_i,b_i)(a_j,b_j)(p_k,q_k)}=\int d^2z_3 d^2 z_4 \sqrt{g}\Psi_{a_ib_i}^{{\cal M}_1 \dagger} \Psi_{a_jb_j}^{{\cal M}_2} \Phi_{p_kq_k}^{{\cal M}_3},
\label{eq:Yukawacop}
\end{eqnarray}
where $\Psi_{ab}^{\cal M}$ and $\Phi_{pq}^{\cal M}$ denote the fermionic and bosonic wavefunctions~(\ref{eq:fw}) and~(\ref{eq:bw}), respectively, each one labeled 
by a pair of integer $(a,b)=(a_i,b_i),(a_j,b_j)$ and $(p,q)=(p_k,q_k)$ characterizing its generation, and a flux number ${\cal M}_1, {\cal M}_2. {\cal M}_3$ defined below. 
The subscripts $i$, $j$ and $k$ indicate a specific generation among those allowed by the normalization and validity conditions for bosons~(\ref{eq:gen}),~(\ref{eq:gen2}) 
and fermions~(\ref{eq:genfer}),~(\ref{eq:genfer2}), respectively.
All of these wavefunctions belong to bifundamental representations under a certain pair among the product subgroups of the original $U(N)$ group broken by the flux. 
Because each bifundamental field feels a difference of two fluxes, the above flux number ${\cal M}$ carrying such information is defined as 
${\cal M}_u\equiv \{ M_1^u,M_2^u\}=\{ M_{1(N_{s_u})}-M_{1(N_{t_u})} ,M_{2(N_{s_u})}-M_{2(N_{t_u})}\}$ for  $s_u,t_u=1,\ldots n$, where $ (s_u,t_u) \neq (s_{u'},t_{u'})$, $u\neq u'$ and $(u,u'=1,2,3)$. 
In the following, we analyze the behavior of overlap integrals~(\ref{eq:Yukawacop}) for some choice of fluxes from a phenomenological point of view, 
although we do not specify a full embedding of all the matter fields into these bifundamental representations. 
(See Refs.~\cite{Abe:2012fj,Abe:2012ya} for a concrete example of the full embedding in the case of factorizable tori.) 

We consider the model with non-supersymmetric fluxes written by Eq.~(\ref{eq:FI}), namely $M_1^s=M_2^s$ with $s=1,2$ and $3$. 
As for the overlap integrals between the boson $\Phi_{24,p'_kq'_k}^{{\cal M}_3}$ and the fermions $\Psi_{3,a_ib_i}^{{\cal M}_1\dagger}$, $\Psi_{3a_jb_j}^{{\cal M}_2}$, the integrand in Eq.~(\ref{eq:Yukawacop}) is expressed as
\begin{eqnarray}
\sqrt{g}\Psi_{3,a_ib_i}^{{\cal M}_1\dagger}\Psi_{3a_jb_j}^{{\cal M}_2}\Phi_{24,p'_kq'_k}^{{\cal M}_3}=J(|z_3|^2, |z_4|^2)\times z_3^{a_i+p'_k}z_4^{b_i+q'_k}\bar{z}_3^{a_j}\bar{z}_4^{b_j},
\label{eq:Yukawaint}
\end{eqnarray}
where $M_1^1, M_1^2>0$, $M_1^3<0$ and $J(|z_3|^2, |z_4|^2)$ denotes the real function. 
In Eq.~(\ref{eq:Yukawaint}), non-vanishing overlap integrals require that they are entirely real valued. Such a requirement is satisfied under
\begin{eqnarray}
a_i+p'_k=a_j, \quad b_i+q'_k=b_j.
\label{eq:ppm}
\end{eqnarray}
Similarly, for the other combination of zero-modes, $\Psi_{3,a_ib_i}^{{\cal M}_1 \dagger} \Psi_{4,a'_jb'_j}^{{\cal M}_2} \Phi_{13,p_kq_k}^{{\cal M}_3}$, 
the overlap integrals with $M_1^1, M_1^3>0$ and $M_1^2<0$ are non-vanishing under the following condition
\begin{eqnarray}
a_i+a'_j=p_k, \quad b_i+b'_j=q_k,
\end{eqnarray}
and so on. 

Then we analyze one particular example and show the values of overlap integrals evaluated by Eq.~(\ref{eq:Yukawacop}) explicitly. Being aware of the observed three generations of SM fermions, in the following, 
we adopt the three fermionic wavefunctions drawn in Fig.~\ref{fig:bs_FI2}, those satisfy validity conditions and are labeled by $i,j=1,2,3$. 
By choosing the several numbers of fluxes with the fixed volume ${\cal V}_1 = {\cal V}_2 \equiv {\cal V}$ of D$7$-brane parametrized by 
\begin{eqnarray}
c=1,\,\,\, L\simeq 13.5,\,\,\, M_s=1.08\times 10^{17}\,{\rm GeV},\nonumber\\
{\cal M}_1{\cal V}={\cal M}_2{\cal V}=\{ 1,1\}, \,\,\,{\cal M}_3{\cal V}_=\{ -2,-2\},
\label{eq:para4}
\end{eqnarray}
the following pair of integers are allowed by Eqs.~(\ref{eq:genfer}) and~(\ref{eq:genfer2}),
\begin{eqnarray}
(a_i,b_i)=(0,0),(0,1),(1,0),\,\,\,(a_j,b_j)=(0,0),(0,1),(1,0),
\end{eqnarray}
yielding three generations for each fermionic zero-mode. 
In this case, there are potentially non-zero overlap integrals between twenty-one generations of boson and three generations of fermions, those are characterized by Eqs.~(\ref{eq:gen}) and~(\ref{eq:gen2}) as
\begin{align}
(p'_k,q'_k)=&(-1,-1),(-1,0),(-1,1),(-1,2),(-1,3),(-1,4),\nonumber\\
&(0,-1),(0,0),(0,1),(0,2),(0,3)\nonumber\\
&(1,-1),(1,0),(1,1),(1,2),\nonumber\\
&(2,-1),(2,0),(2,1)\nonumber\\
&(3,-1),(3,0),\nonumber\\
&(4,-1).
\end{align}
Among them, we find the overlap integrals for fourteen pairs of integers vanish. The remaining seven pairs,
\begin{eqnarray}
(p'_k,q'_k)=(0,0),(0,1),(1,0),(-1,0),(-1,1),(0,-1),(1,-1),
\label{eq:cf}
\end{eqnarray}
have a possibility to yield non-vanishing overlap integrals. 
Hereafter the seven generation numbers $k=0,1,2,3,4,5,6$ ($i,j=0,1,2$) indicate (the first three of) the seven pairs of integers~(\ref{eq:cf}), respectively,
e.g., $Y_{203}=Y_{(1,0)(0,0)(-1,0)}$ for $Y_{ijk}$ with $i=2$, $j=0$ and $k=3$. 

Assuming non-vanishing vacuum expectation values (VEVs) for bosons $\Phi_{24,p'_kq'_k}^{{\cal M}_3}$ denoted by $\langle H_k\rangle$, 
we find the fermion masses are given by the eigenvalues of the following mass matrices, 
\begin{eqnarray}
M_{ij}=\sum_{k=0}^6 Y_{ijk} \langle H_k\rangle,
\end{eqnarray}
where
\begin{eqnarray}
Y_{ij0}=
\left(
\begin{array}{ccc}
Y_{000}&$0$&$0$\\
$0$&Y_{110}&$0$\\
$0$&$0$&Y_{220}
\end{array}
\right) ,\quad
Y_{ij1}=
\left(
\begin{array}{ccc}
$0$&Y_{011}&$0$\\
$0$&$0$&$0$\\
$0$&$0$&$0$
\end{array}
\right) ,\quad
Y_{ij2}=
\left(
\begin{array}{ccc}
$0$&$0$&Y_{022}\\
$0$&$0$&$0$\\
$0$&$0$&$0$
\end{array}
\right) ,\nonumber\\
Y_{ij3}=
\left(
\begin{array}{ccc}
$0$&$0$&$0$\\
$0$&$0$&$0$\\
Y_{203}&$0$&$0$
\end{array}
\right) ,\quad
Y_{ij4}=
\left(
\begin{array}{ccc}
$0$&$0$&$0$\\
$0$&$0$&$0$\\
$0$&Y_{214}&$0$
\end{array}
\right) ,\nonumber\\
Y_{ij5}=
\left(
\begin{array}{ccc}
$0$&$0$&$0$\\
Y_{105}&$0$&$0$\\
$0$&$0$&$0$
\end{array}
\right) ,\quad
Y_{ij6}=
\left(
\begin{array}{ccc}
$0$&$0$&$0$\\
$0$&$0$&Y_{126}\\
$0$&$0$&$0$
\end{array}
\right) .
\end{eqnarray}
For example, the following ratios of bosonic VEVs, 
\begin{eqnarray}
\frac{\langle H_k\rangle}{\langle H_0\rangle} =(1,9,7,1,7,1,8),
\end{eqnarray}
lead to fermion mass ratios  $(m_1,m_2,m_3)/m_3=(1.13\times 10^{-3},2.20\times 10^{-2},1)$ which have the similar structure for down-type quark mass ratios compared with the observed ones~\cite{Agashe:2014kda} $(m_d,m_s,m_b)$ $/m_b=(1.17\times 10^{-3},2.27\times 10^{-2},1)$. 
As we see in this simple example, the analytic form of wavefunctions and their overlap integrals 
are important tools for constructing particle physics models involving D$7$-branes on the conifold. We leave such model building for future works.

\subsection{The possible embeddings of conifold into the global CY} 
\label{subsec:global}
In earlier discussions, when the global CY manifold is described by 
the conifold metric, the cycle of D$7$-brane involved in the conifold 
has an infinite volume caused by the noncompact radial direction 
of $AdS_5$, $r \in [0,\infty]$. According to it, the magnetic flux possessed by such a D$7$-brane (\ref{eq:flux}) diverges. 
Note that it is only valid in near-horizon limit in Klebanov-Witten background. 
In order to avoid this problem, it is significant to consider the 
local description of the conical geometry in the global compact CY. 

Here we briefly review the recent development of constructing such a compact CY. 
One of the method for the construction is proposed by Batyrev in  Ref.~\cite{Batyrev:1994hm} by employing the toric variety, 
where the compact Calabi-Yau is embedded as a hypersurface in the ambient complex four-dimensional 
compact toric variety defined by a reflexive polytope 
in the terminology of toric construction.
Especially, such reflexive polytopes which include the compact Calabi-Yau threefolds is systematically found by 
Kreuzer and Skarke~\cite{Kreuzer:2000xy}, where the number of them are estimated as $473800776$ using a computational code. 
The software package known as PALP \cite{Kreuzer:2002uu} was released and 
the list of the detail data can be found on the web page~\cite{CYurl}.
See a review in Ref.~\cite{Knapp:2011ip} for a more detailed discussion. 

It is expected that the local configuration considered in this paper is  achieved as a certain limit of global CY. 
Though in a different context, the flux configurations on D$7$-branes in $4$D toric variety is discussed in Ref.~\cite{Balasubramanian:2012wd,Cicoli:2012vw}. 

\section{Conclusion}
\label{sec:conclu}
In this paper, we have studied the zero-mode wavefunctions 
on local D$7$-brane with and without a magnetic flux 
in the $AdS_5 \times T^{1,1}$ background 
where the large number of D$3$-branes are placed at the tip of conifold, so called the Klebanov-Witten model~\cite{Klebanov:1998hh}. 
We have considered the case that the D$7$-branes wrap the internal cycles in 
the conifold in a kappa-symmetric way, which ensures the 
stability of D$7$-brane. 
The kappa-symmetry condition determines the induced metric of such D$7$-brane. 
We have explored the zero-mode wavefunctions by solving the Laplace and Dirac equations employing the explicit metric.

If some magnetic fluxes are turned on in the internal cycles wrapped by D$7$-branes, the charged fields in general have degenerate zero-modes~\cite{Cremades:2004wa} identified with the generation of matter fields living on the D$7$-brane. 
We have found that they are exponentially localized around the tip of conifold due to the nature of dilogarithm function. 

Especially, in view of the 5D effective theory, the wavefunctions of matter fields in $AdS_5$ have the exponential form depending on the number of magnetic fluxes they feel.
In Sec.~\ref{sec:AdS5}, we have analyzed the detailed form of wavefunctions localized near the tip of conifold due
to the magnetic fluxes, whereas the graviton localizes towards the CY manifold.
Such a situation can yield various small mass scales compared with the Planck scale when the localized fields obtain non-vanishing vacuum expectation values~\cite{Randall:1999ee}, which is quite interesting from the viewpoint of phenomenological model building.
It is remarkable that, in the terminology of $AdS_5$ supergravity~\cite{Altendorfer:2000rr} compactified on $S^1/Z_2$, the so-called bulk mass~\cite{Chang:1999nh} (proportional to a graviphoton charge) of matter fields is determined effectively, in our conifold setup, by the magnetic fluxes. Note also that the bulk mass is one of the key parameters in the particle-physics model building on $AdS_5$.\footnote{The exponential localization of fields are quite useful for phenomenological/cosmological model building
in (a slice of) $AdS_5$. For example, Ref.~\cite{Otsuka:2015oma} addressed most phenomenological and cosmological issues in the
modern particle physics in a single model based on the $AdS_5$ supergravity.}
%We have shown that the down-type quark mass ratio can be generated with appropriate choices of flux configurations and down-type Higgs VEV ratios. 

We have assumed, throughout this paper, that the conifold background could be glued to a certain global CY manifold 
in order to determine the compact cycle of D$7$-brane in terms of the global description. 
We expect that the resultant wavefunctions do not receive sizable corrections from the global CY manifold. 
This assumption seems to be appropriate since we have obtained wavefunctions for some flux configurations satisfying the validity conditions, 
which extract profiles localized toward the tip of conifold and converging to zero in the direction of global CY. 

It is important to study the global embedding of our local results by employing the Batyrev's method and the Kreuzer-Skarke list mentioned in Sec.~\ref{subsec:global}, 
because such a embedding allows us to identify the exact cycles of D-branes along the direction of the global CY. 
An extension to the warped deformed and/or resolved conifold is also interesting. 
There is a work studying supersymmetric D$7$-branes on the warped deformed conifold 
taking kappa-symmetry condition into account with H-flux \cite{Chen:2008jj}. 
On the other hand, extensions to the geometry with the other types of Sasaki-Einstein manifold, such as $AdS_5\times Y^{p,q}$ and $AdS_5 \times L^{a,b,c}$, 
can be tried referring to a kappa-symmetric embedding \cite{Canoura:2005uz,Canoura:2006es} (see also \cite{Fernandez:2008vh} for a comprehensive review).

Although our analyses in this paper is mainly motivated by the case that D$7$-branes yield (a part of) SM fields, as discussed in Ref.~\cite{Cascales:2003wn}, a semi-realistic spectrum can be 
realized on D$3$-branes at the orbifold singularity at the bottom of a 
warped throat such as the conifold.
These D-brane configurations are also interesting from the 
cosmological point of view in a model building on the conifold. The additional D$7$-branes play important roles in the context of K\"{a}hler moduli stabilization. 
On these warped backgrounds, it has been 
pointed out a possibility of brane inflation on the warped deformed conifold~\cite{Kachru:2003sx} 
as well as a natural inflation on the warped resolved conifold~\cite{Kenton:2014gma}. 
Our results for general fields on D$7$-branes will be applicable to various phenomenological/cosmological scenarios on the conifold including them. We will study these issues elsewhere.

\subsection*{Acknowledgement}
The work of H. A. was supported in part by the Grant-in-Aid for Scientific Research No. 25800158 from the Ministry of Education, Culture, Sports, Science and Technology (MEXT) in Japan. H. O. was supported in part by the Grant-in-Aid for JSPS Fellows No. 26-7296.

\appendix
\section{Kappa-symmetry}
\label{sec:A}
The Killing spinor on $AdS_5$ is given by 
\begin{eqnarray}
\epsilon =r^{\frac{\Gamma_*}{2}}\left( 1+\frac{\Gamma_r}{2L^2}x^\alpha \Gamma_{x^\alpha}(1-\Gamma_*) \right) \eta,
\end{eqnarray}
with
\begin{eqnarray}
\Gamma_* =i\Gamma_{x^0x^1x^2x^3}, 
\end{eqnarray}
and the stable Dp-branes on the background $AdS_5 \times T^{1,1}$ are properly embedded in the kappa-symmetric way. 
As stated in Ref.~\cite{Lu:1996rhb}, the $\kappa$-symmetric conditions 
are equivalent to the following condition,
\begin{eqnarray}
\Gamma_\kappa \epsilon =\epsilon,
\label{eq:kappasym}
\end{eqnarray}
where 
\begin{eqnarray}
\Gamma_\kappa =\frac{1}{(p+1)!\sqrt{-g}}\epsilon^{\mu_1 \cdots \mu_{p+1}}(\tau_3)^{\frac{p-3}{2}}i\tau_2 \otimes \gamma_{\mu_1 \cdots \mu_{p+1}},
\label{eq:Kappap}
\end{eqnarray}
with the pull-back $\tilde{\gamma}_{\mu_1 \cdots \mu_{p+1}}$ of the Gamma matrices. The D$7$-brane considered in this paper satisfies the above condition~(\ref{eq:kappasym}) 
which is summarized in Ref.~\cite{Arean:2004mm}.

\section{Spin connections}
\label{sec:B}
By solving Cartan structure equations,
\begin{eqnarray}
de^a+w^a_b \wedge e^b =0,
\end{eqnarray}
we obtain the nonvanishing spin-connections as follows: 
\begin{align}
&w^{12}_{z_3}=i\frac{\bar{z}_3}{6\rho |z_3|^2(|z_3|^2+c^2)} \left[ -2|z_3|^4+|z_3|^2|z_4|^2-3c^2|z_4|^2 \right], \,\,
w^{12}_{\bar{z}_3}=\overline{w^{12}_{z_3}}, 
\nonumber\\
&w^{12}_{z_4}=i\frac{\bar{z}_4}{6\rho |z_4|^2(|z_4|^2+c^2)} \left[ -2|z_4|^4+|z_3|^2|z_4|^2-3c^2|z_3|^2 \right], \,\,
w^{12}_{\bar{z}_4}=\overline{w^{12}_{z_4}},
\nonumber\\
&w^{13}_{z_3}=\frac{c|z_4|\bar{z}_3}{2\rho \sqrt{6\rho}|z_3|(|z_3|^2+c^2)\sqrt{\frac{1}{9}+\frac{c^2}{6\rho}}}\left[ |z_3|^2 +c^2 \right],\,\,
w^{13}_{\bar{z}_3}=\overline{w^{13}_{z_3}},
\nonumber\\
&w^{13}_{z_4}=-\frac{c|z_3|\bar{z}_4}{2\rho \sqrt{6\rho}|z_4|(|z_4|^2+c^2)\sqrt{\frac{1}{9}+\frac{c^2}{6\rho}}}\left[ |z_4|^2 +c^2 \right],\,\,
w^{13}_{\bar{z}_4}=\overline{w^{13}_{z_4}},
\nonumber\\
&w^{14}_{z_3}=i\frac{c|z_4|\bar{z}_3}{6\rho \sqrt{6\rho}|z_3|(|z_3|^2+c^2)\sqrt{\frac{1}{9}+\frac{c^2}{6\rho}}}\left[ -2\rho +3(|z_3|^2-c^2)\right],\,\, 
w^{14}_{\bar{z}_3}=\overline{w^{14}_{z_3}},
\nonumber\\
&w^{14}_{z_4}=-i\frac{c|z_3|\bar{z}_4}{6\rho \sqrt{6\rho}|z_4|(|z_4|^2+c^2)\sqrt{\frac{1}{9}+\frac{c^2}{6\rho}}}\left[ -2\rho +3(|z_4|^2-c^2)\right],\,\,
w^{14}_{\bar{z}_4}=w^{14}_{z_4},
\nonumber
\end{align}
\begin{align}
&w^{23}_{z_3}=-i\frac{c|z_4|\bar{z}_3}{2\rho \sqrt{6\rho}|z_3|(|z_3|^2+c^2)\sqrt{\frac{1}{9}+\frac{c^2}{6\rho}}}\left[ |z_3|^2-c^2 \right],\,\,
w^{23}_{\bar{z}_3}=\overline{w^{23}_{z_3}},
\nonumber\\
&w^{23}_{z_4}=i\frac{c|z_3|\bar{z}_4}{2\rho \sqrt{6\rho}|z_4|(|z_4|^2+c^2)\sqrt{\frac{1}{9}+\frac{c^2}{6\rho}}}\left[ |z_4|^2-c^2 \right],\,\,
w^{23}_{\bar{z}_4}=\overline{w^{23}_{z_4}},
\nonumber\\
&w^{24}_{z_3}=\frac{c|z_4|\bar{z}_3}{6\rho \sqrt{6\rho}|z_3|(|z_3|^2+c^2)\sqrt{\frac{1}{9}+\frac{c^2}{6\rho}}}\left[ 2\rho +3(|z_3|^2+c^2) \right],\,\,
w^{24}_{\bar{z}_3}=\overline{w^{24}_{z_3}},
\nonumber\\
&w^{24}_{z_4}=-\frac{c|z_3|\bar{z}_4}{6\rho \sqrt{6\rho}|z_4|(|z_4|^2+c^2)\sqrt{\frac{1}{9}+\frac{c^2}{6\rho}}}\left[ 2\rho +3(|z_4|^2+c^2) \right],\,\,
w^{24}_{\bar{z}_4}=\overline{w^{24}_{z_4}},
\nonumber\\
&w^{34}_{z_3}=-i\frac{c^2\bar{z}_3}{36\rho^2 (|z_3|^2+c^2)\left( \frac{1}{9}+\frac{c^2}{6\rho} \right)}\left[ \rho +3(|z_4|^2+c^2) \right],\,\,
w^{34}_{\bar{z}_3}=\overline{w^{34}_{z_3}},
\nonumber\\
&w^{34}_{z_4}=-i\frac{c^2\bar{z}_4}{36\rho^2 (|z_4|^2+c^2)\left( \frac{1}{9}+\frac{c^2}{6\rho} \right)}\left[ \rho +3(|z_3|^2+c^2) \right],\,\,
w^{34}_{\bar{z}_4}=\overline{w^{34}_{z_4}}.
\end{align}
%\section{Wavefunctions on the D5-brane}

\end{document}